\documentclass[hyper,letterpaper]{JHEP3}

\usepackage{epsfig}
\usepackage{amsbsy}
\usepackage{varioref} 
\usepackage{pifont}
\usepackage{amsmath} 
\usepackage{graphicx}
\usepackage{axodraw}

\newcommand{\tr}{\mbox{tr}}

\def\slashii#1{\setbox0=\hbox{$#1$}             
   \dimen0=\wd0                                 
   \setbox1=\hbox{\sl/} \dimen1=\wd1            
   \ifdim\dimen0>\dimen1                        
      \rlap{\hbox to \dimen0{\hfil\sl/\hfil}}   
      #1                                        
   \else                                        
      \rlap{\hbox to \dimen1{\hfil$#1$\hfil}}   
      \hbox{\sl/}                               
   \fi}                                         %
\def\slashiii#1{\setbox0=\hbox{$#1$}#1\hskip-\wd0\hbox to\wd0{\hss\sl/\/\hss}}
\title{The Three Site Model at One-Loop}

\author{R. Sekhar Chivukula and Elizabeth H. Simmons\\
Department of Physics and Astronomy, Michigan State University\\
East Lansing, MI 48824, USA\\
E-mail: sekhar@msu.edu, esimmons@msu.edu}

\author{
Shinya Matsuzaki\\
Department of Physics, Nagoya University\\
Nagoya 464-8602, Japan\\
	E-mail: synya@eken.phys.nagoya-u.ac.jp}

\author{
Masaharu Tanabashi\\
Department of Physics, Tohoku University\\
Sendai 980-8578, Japan\\
	E-mail: tanabash@tuhep.phys.tohoku.ac.jp}

\abstract{
In this paper we compute  the one-loop chiral logarithmic corrections to
all ${\cal O}(p^4)$ counterterms in the three site Higgsless model.
The calculation is performed using the background
field method for both the chiral- and gauge-fields, and using Landau gauge
for the quantum fluctuations of the gauge fields. The results agree with our previous 
calculations of the chiral-logarithmic corrections to the $S$ and $T$
parameters in 't Hooft-Feynman gauge.  The work reported here
includes a complete evaluation of all one-loop divergences in an $SU(2) \times U(1)$
nonlinear sigma model, corresponding to an electroweak effective Lagrangian
in the absence of custodial symmetry.\\ \\
\centerline{March 7, 2007}}


\preprint{MSUHEP-070221\\
TU-784}

\begin{document}

\section{Introduction}

Higgsless models
\cite{Csaki:2003dt} achieve electroweak
symmetry breaking without introducing a fundamental scalar Higgs
boson \cite{Higgs:1964ia}, and the unitarity of longitudinally-polarized
$W$ and $Z$ boson scattering  is preserved by the exchange of extra vector
bosons \cite{SekharChivukula:2001hz,Chivukula:2002ej,Chivukula:2003kq,He:2004zr}.
Inspired by TeV-scale \cite{Antoniadis:1990ew} compactified five-dimensional
gauge theories  \cite{Agashe:2003zs,Csaki:2003zu,Burdman:2003ya,Cacciapaglia:2004jz}, these models provide effectively unitary descriptions of the electroweak sector beyond the TeV energy scale. 

Five-dimensional gauge-theories are not renormalizable, and therefore Higgsless models
can only be viewed as effective theories valid below some high-energy cutoff (above which
some other physics, a ``high-energy" completion, must be present). Since these theories
are only low-energy effective theories their properties may be conveniently studied
using deconstruction \cite{Arkani-Hamed:2001ca,Hill:2000mu}, which is a technique
to build a four-dimensional gauge theory, with an appropriate gauge-symmetry breaking pattern,
which approximates the properties of a five-dimensional theory. Deconstructed Higgsless models \cite{Foadi:2003xa,Hirn:2004ze,Casalbuoni:2004id,Chivukula:2004pk,Perelstein:2004sc,Georgi:2004iy,SekharChivukula:2004mu} have been used as tools to compute the general properties of Higgsless theories, and to illustrate the phenomological properties of this class of models. 

The simplest deconstructed Higgsless model \cite{SekharChivukula:2006cg} incorporates only three
sites on the deconstructed lattice, and the only additional vector states (other than the
usual electroweak gauge bosons) are a triplet of $\rho^{\pm}$ and $\rho^0$ mesons (which
may be interpreted as the lightest Kaluza-Klein states of a compactified five-dimensional
theory).
This theory is in the same class as models of extended  electroweak gauge symmetries \cite{Casalbuoni:1985kq,Casalbuoni:1996qt}  motivated by models of hidden local symmetry \cite{Bando:1985ej,Bando:1985rf,Bando:1988ym,Bando:1988br,Harada:2003jx} in QCD, and the gauge sector is precisely that of the BESS model \cite{Casalbuoni:1985kq}. While simple, the three site model  is
sufficiently rich \cite{SekharChivukula:2006cg} to describe the physics associated with fermion mass generation,
as well as the  fermion delocalization \cite{Cacciapaglia:2004rb,Cacciapaglia:2005pa,Foadi:2004ps,Foadi:2005hz,Chivukula:2005bn,Casalbuoni:2005rs,SekharChivukula:2005xm}
required\footnote{It should be emphasized, however, that there is no explanation
in any of these models (which are only low-energy effective theories)
for the amount of delocalization. In particular, there is no 
dynamical reason why the fermion delocalization present
{\it must} be such as to make the value of $\alpha S$
small.} in order to accord with precision electroweak tests
\cite{Peskin:1992sw,Altarelli:1990zd,Altarelli:1991fk,Barbieri:2004qk,Chivukula:2004af}.

Recently, we have computed \cite{Matsuzaki:2006wn}  the one-loop chiral logarithmic corrections to the $S$ and $T$ parameters \cite{Peskin:1992sw,Altarelli:1990zd,Altarelli:1991fk} in the three site Higgsless model, in the limit $M_W \ll M_{\rho} \ll \Lambda$, where $\Lambda$ is the cutoff of the effective three site Higgsless theory. In ref. \cite{Matsuzaki:2006wn}, the 
calculation was performed by directly computing the
one-loop corrections to four-fermion scattering processes in 't~Hooft-Feynman gauge, including the ghost, unphysical Goldstone-boson, and appropriate  ``pinch" contributions \cite{Degrassi:1992ue,Degrassi:1992ff} required to obtain gauge-invariant results for the one-loop self-energy functions. 

In this paper, we compute the one-loop
corrections to {\it all} ${\cal O}(p^4)$ counterterms in the three site Higgsless model, using
the renormalization group equation (RGE) technique. 
The calculation here is performed using the background
field method for both the chiral- and gauge-fields, and using Landau gauge
for the quantum fluctuations of the gauge fields. Focusing on those corrections
which contribute to $S$ and $T$, we find that our RGE results
agree with the chiral-logarithmic corrections to the $S$ and $T$
parameters determined previously \cite{Matsuzaki:2006wn} in 't Hooft-Feynman gauge,
thereby establishing the gauge-invariance of our results directly.\footnote{Recently,
$\alpha S$ and $\alpha T$ have been computed in a three site model with linear sigma-model
link fields \protect\cite{Tomohiro}. In the limit in which the extra scalars in this model
are heavier than the vector-bosons, the leading-log contributions agree with
the results of \protect\cite{Matsuzaki:2006wn}, providing another check of those
calcualtioins.} Our calculations
include also a complete evaluation of all one-loop divergences in an $SU(2) \times U(1)$
nonlinear sigma model, corresponding to an electroweak effective Lagrangian
in the absence of custodial symmetry.

The hierarchy of scales, 
$M_W \ll M_{\rho} \ll \Lambda$, implies that the calculation breaks up into
two different energy regimes. For renormalization group scales $\mu$ such that
$M_\rho < \mu < \Lambda$, we consider the running of the effective Lagrangian
parameters in the full three site model (illustrated in Figure~\ref{fig:threesite}). Running in
the three site model is discussed in Section \ref{sec:threesite}. At an energy
scale of order $M_\rho$, one must ``integrate out" the $\rho$-meson, matching to an effective
two site model (the electroweak chiral Lagrangian \cite{Appelquist:1980ae,Appelquist:1980vg,Longhitano:1980iz,Longhitano:1980tm,Appelquist:1993ka}, illustrated in Figure~\ref{fig:twosite2ew}). For renormalization group scales $\mu<M_\rho$, we 
consider the running of the 
parameters in the two site electroweak chiral Lagrangian. Both the matching of the
three site to the two site model and the low-energy running in the two site model are discussed
in Section~\ref{sec:twosite}. In Section~\ref{sec:st}, we discuss the chiral-logarithmic corrections
to the $S$ and $T$ parameters, and discuss the correspondence of the RGE calculation
with that presented in Ref.~\cite{Matsuzaki:2006wn}. Section 5 contains a concluding summary
of our results.

The appendices include a detailed description of the one-loop calculations leading to the RGE equations. The one-loop renormalization of the operators \cite{Appelquist:1980ae,Appelquist:1980vg,Longhitano:1980iz,Longhitano:1980tm,Appelquist:1993ka} in the $SU(2)\times SU(2) / SU(2)$ chiral Lagrangian was first undertaken in Refs.  \cite{Weinberg:1978kz,Gasser:1983yg,Gasser:1984gg} and later discussed in the context of electroweak physics \cite{Herrero:1993nc,Dobado:1997jx}.  We review the  the RGE calculation in a gauged $SU(2) \times SU(2)$ model using the background field method for
the chiral- and gauge-fields,   and add a calculation of the renormalization of the operator associated with fermion delocalization effects.  Then, in the same language, we  report our RGE calculation for a gauged $SU(2) \times U(1)$ model.  We then show that the results of these separate calculations may be combined to obtain  the RGE equations for the three site gauged $SU(2) \times SU(2) \times U(1)$ model.

\section{The Three Site Model}

\label{sec:threesite}

\EPSFIGURE[tb]
 {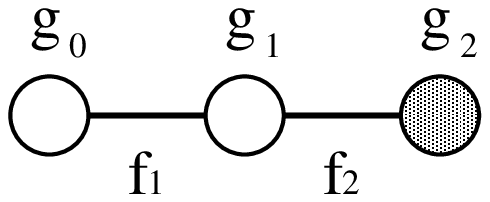,width=5cm}
 {The moose diagram \protect\cite{Georgi:1985hf} for the three site model analyzed in this
  note. The $SU(2)$ gauge groups are shown as open circles, and the $U(1)$ gauge
  group is shown as a shaded circle.  The fermion $SU(2)$ couplings arise from sites 0 
  and 1, and their hypercharge couplings from site 2 ({\it c.f.} eqn.~(\protect\ref{eq:fermion_coupling})). The links, read from left to
  right, correspond to nonlinear sigma model fields $U_1$ and $U_2$ with $f$-constants
  $f_1$ and $f_2$, respectively. \label{fig:threesite} }

In this section we consider the renormalization group structure of the
three site model shown in Figure~\ref{fig:threesite}, valid for energies below
the cutoff $\Lambda$ but above the mass of the heavy vector meson $M_\rho$.

\subsection{The three site Lagrangian and counterterms}

The lowest order (${\cal O}(p^2)$) custodially symmetric Lagrangian of this model is given by
\begin{equation}
  {\cal L}_2 = 
    \sum_{i=1}^{2} \dfrac{f_i^2}{4} 
    \tr\left[(D_\mu U_i)^\dagger (D^\mu U_i) \right]
   -\sum_{i=0}^{2} \dfrac{1}{2g_i^2}
    \tr\left[V_{i\mu\nu}V_i^{\mu\nu} \right],
\label{eq:p2}
\end{equation}
where $D_\mu U_i$ and $V_{i\mu\nu}$ are defined as
\begin{equation}
  D_\mu U_i \equiv \partial_\mu U_i + i V_{(i-1)\mu} U_i
             - i U_i V_{i\mu},
\end{equation}
\begin{equation}
  V_{i\mu\nu} \equiv \partial_\mu V_{i\nu} - \partial_\nu V_{i\mu}
               +i[V_{i\mu}, V_{i\nu}].
\end{equation}
The first two gauge fields ($i=0,1$) correspond to $SU(2)$ gauge groups,
\begin{equation}
  V_{i\mu} = \sum_{a=1,2,3} \frac{\tau^a}{2} V^a_{i\mu}, \qquad
  \mbox{for $i=0,1$},
\end{equation}
while the last gauge field ($i=2$) corresponds to $U(1)$ group
embedded as the $T_3$ generator of $SU(2)$,
\begin{equation}
  V_{2\mu} = \frac{\tau^3}{2} V^3_{2\mu}, \qquad
  \mbox{for $i=2$}. 
\end{equation}
There is one additional ${\cal O}(p^2)$ term violating custodial
symmetry, 
\begin{equation}
  {\cal L}_2' = \beta_{(2)} \dfrac{f_2^2}{4} 
    \tr\left[ U_2^\dagger (D_\mu U_2)\tau^3 \right]
    \tr\left[ U_2^\dagger (D^\mu U_2)\tau^3 \right].
\label{eq:beta2}
\end{equation}

Fermions (quarks and leptons) couple to sites 0 and 1 (weak isospin),
and to site 2 (weak hypercharge) through,
\begin{equation}
  {\cal L}_f = -2J_W^{a\mu}\,
                \tr\left[\frac{\tau^a}{2} V_{0\mu}\right]
               -2x_1 J_W^{a\mu}\,
                \tr\left[ \frac{\tau^a}{2} i (D_\mu U_1) U_1^\dagger \right]
               -2J_Y^{\mu} \,
                \tr\left[\frac{\tau^3}{2} V_{2\mu}\right],   
\label{eq:fermion_coupling}
\end{equation}
where
\begin{equation}
  J_W^{a\mu} \equiv \bar{\psi}_L \gamma^\mu T^a \psi_L, \qquad
  J_Y^{\mu} \equiv \bar{\psi}_L \gamma^\mu Y_{\psi_L} \psi_L 
                  +\bar{\psi}_R \gamma^\mu Y_{\psi_R} \psi_R,
\end{equation}
and the $T^a$ and $Y_{\psi_L,\psi_R}$ are the $SU(2)_W$ and $U(1)_Y$ charges
of the fermions.

In order to renormalize the one-loop divergences of this model, we need to
introduce appropriate counter terms at ${\cal O}(p^4)$,
\begin{equation}
  {\cal L}_4 = \sum_{i=1}^2 {\cal L}_{4}^{(i)},
\end{equation}
with
\begin{eqnarray}
  {\cal L}_{4}^{(i)} 
  &=& \alpha_{(i)1} \tr\left[ 
        V_{(i-1)\mu\nu} U_i V_{i}^{\mu\nu} U_i^\dagger
      \right]
  \nonumber\\
  & & -2i\alpha_{(i)2} \tr\left[
        (D_\mu U_i)^\dagger (D_\nu U_i) V_{i}^{\mu\nu}
      \right]
  \nonumber\\
  & & -2i\alpha_{(i)3} \tr\left[
        V_{(i-1)}^{\mu\nu} (D_\mu U_i) (D_\nu U_i)^\dagger
      \right] 
  \nonumber\\
  & & +\alpha_{(i)4} 
       \tr\left[(D_\mu U_i) (D_\nu U_i)^\dagger \right] 
       \tr\left[(D^\mu U_i) (D^\nu U_i)^\dagger \right] 
  \nonumber\\
  & & +\alpha_{(i)5} 
       \tr\left[(D_\mu U_i) (D^\mu U_i)^\dagger \right] 
       \tr\left[(D_\nu U_i) (D^\nu U_i)^\dagger \right].
       \label{eq:twoten}
\end{eqnarray}
Here we have neglected terms violating the custodial symmetry
at ${\cal O}(p^4)$ level. Of the terms introduced in (\ref{eq:beta2}) and (\ref{eq:twoten}), the $\alpha_{(i)1}$ and $\beta_{(2)}$ will be
of particular interest, as they contribute, respectively, to the precision electroweak parameters
$\alpha S$ and $\alpha T$.

The three site model approximates (see \cite{SekharChivukula:2006cg} for
details) the standard electroweak theory in the limit
\begin{equation} 
g_0 \ll g_1~,\ \  \ \ \ \ \ g_2 \ll g_1~,
\end{equation}
and we therefore define a small parameter $x = g_0/g_1 \ll 1$.
For simplicity, we also take $f_1=f_2$.
The analysis of the three site model proceeds in an expansion in
powers of $x$, and at tree-level the values of all electroweak observables
differ from those in the standard model beginning at ${\cal O}(x^2)$.  To leading order,
we find that $g_{0,2}$ are approximately equal to the standard model $SU(2)_W$ and
$U(1)_Y$ couplings. Defining an angle $\theta$ such that $s\equiv\sin\theta$,
$c\equiv\cos\theta$, and $s/c = g_2/g_0$, we find
\begin{equation}
g^2_0 \approx \frac{4\pi \alpha}{s^2} = \frac{e^2}{s^2}~,\ \ g^2_2\approx
\frac{4\pi\alpha}{s^2} = \frac{e^2}{c^2}~,
\end{equation}
where $\alpha$ is the fine-structure constant and $e$ is the charge of the electron.

\subsection{The Size of Electroweak and Radiative Corrections}

To leading order, we find
\begin{equation}
\frac{M^2_W}{M^2_\rho} \approx \frac{x^2}{4}~,
\end{equation}
and \cite{SekharChivukula:2006cg,Matsuzaki:2006wn,Anichini:1994xx} 
that the tree-level values
\begin{eqnarray}
\alpha S^{tree} & = & \frac{4 \pi \alpha}{g^2_1} \left(1- \frac{2x_1}{x^2}\right)\\
& = & \frac{4 s^2 M^2_W}{M^2_\rho}\left(1-\frac{x_1M^2_\rho}{2M^2_W}\right)~,\label{eq:smasseqn}\\
\alpha T^{tree} & = & \beta_{(2)}~,
\end{eqnarray}
summarize the deviations in the three site model relative to the standard electroweak theory.

Observationally, $\alpha S,\,\alpha T \le {\cal O}(10^{-3})$ \cite{Barbieri:2004qk}.
The mass of $M_\rho$ is bounded by ${\cal O}(1\,{\rm TeV})$, since $\rho$-exchange
is necessary to maintain the unitarity of longitudinally polarized $W$-boson scattering --
leading (from eqn.~(\ref{eq:smasseqn})) to a value of $\alpha S$ which is too large  \cite{Chivukula:2004pk,Chivukula:2004af} for
localized fermions with $x_1=0$.  The
phenomenologically preferred region therefore has $x_1 \approx x^2/2$,which is the
condition for ``ideal delocalization" \cite{SekharChivukula:2005xm,Anichini:1994xx} in this model. 
In what follows, we will assume $x_1 = {\cal O}(x^2)$. 

As the observed limits on $\alpha S$ are ${\cal O}(10^{-3})$, radiative electroweak
corrections are potentially important. In the context of the hierarchy $M_W \ll M_\rho \ll \Lambda$,
the leading chiral logarithmic corrections are found to be of order \cite{Matsuzaki:2006wn} 
\begin{equation}
\frac{\alpha}{4\pi} \log\frac{\Lambda}{M_\rho}~, \ \ \ 
\frac{\alpha}{4\pi} \log\frac{M_\rho}{M_{H,{\rm ref}}}~, \ \ 
{\rm or} \ \  \ \ \frac{\alpha\, x_1}{4\pi x^2} \log\frac{\Lambda}{M_\rho}~,
\end{equation}
where $M_{H,ref}$ is the reference Higgs mass used in the extraction
of the value of $\alpha S$ from electroweak observations.
As we shall see, the RGE calculations described here will allow us to reproduce
the chiral log corrections found in Ref. \cite{Matsuzaki:2006wn}, while simultaneously
calculating the corrections to the other ${\cal O}(p^4)$ chiral parameters. 
Given that $\alpha T$
is bounded by ${\cal O}(10^{-3})$, we will assume that the theory is approximately
custodially symmetric and $\beta_{(2)} = {\cal O}(\alpha/4\pi)$. In our computations, therefore,
we will neglect contributions of order $\alpha \beta_{(2)}/4\pi$.

Inspired by AdS/CFT duality \cite{Maldacena:1998re,Gubser:1998bc,Witten:1998qj,Aharony:1999ti}, 
tree-level computations in this
theory are interpreted to represent the leading terms in a large-$N$ expansion \cite{'tHooft:1973jz}
of the strongly-coupled dual gauge theory akin to ``walking technicolor" \cite{Holdom:1981rm,Holdom:1985sk,Yamawaki:1986zg,Appelquist:1986an,Appelquist:1987tr,Appelquist:1987fc}.  Formally,
both the electroweak  \cite{Burdman:2003ya,Matsuzaki:2006wn} and the
chiral corrections \cite{Manohar:1998xv,Chivukula:1992gi} are suppressed by $1/N$, and therefore
the calculations presented here are consistent with duality.  In this language, our discussion of the $S$ and $T$ parameters in the three site model will include (a) the tree-level contributions, which are of order $N$, (b) the chiral corrections which are order $1$ but enhanced by chiral logs, and (c) the effects of the $p^4$ counterterms, which are simply order $1$.

\subsection{RGE Solutions in the three site model}

\label{sec:threesiterunning}

Performing  $\overline{\rm MS}$ renormalization of the one-loop amplitudes, 
we find that the chiral parameters (including, in particular, 
 $\alpha_{(i)1} - \alpha_{(i)5}$ and $\beta_{(2)}$) depend on the
renormalization scale $\mu$. The invariance of the amplitudes for physical
processes with
respect to changes in the renormalization scale $\mu$ gives rise to
renormalization group equations (RGEs) for these chiral parameters in the usual manner.
A detailed description of the calculation
of the renormalization group equations,
which is performed using the background field method for both the chiral-
and gauge-fields and using Landau gauge for the quantum fluctuations of
the gauge field, is given in appendix \ref{sec:rgeappendix} of this paper. 
Here, we simply list the one-loop RGEs which result, for ${\cal O}(p^2)$ parameters,
\begin{eqnarray}
  \mu \dfrac{d}{d\mu} f_1^2 
    &=& \dfrac{3}{(4\pi)^2} (g_0^2 + g_1^2) f_1^2, 
    \label{eq:threesitefbegin}
  \\
  \mu \dfrac{d}{d\mu} f_2^2 
    &=& \dfrac{3}{(4\pi)^2} (g_1^2 + \frac{1}{2} g_2^2) f_2^2, 
  \\
  \mu \dfrac{d}{d\mu} \left( \beta_{(2)} f_2^2 \right) 
    &=& \dfrac{3}{4(4\pi)^2} g_2^2 f_2^2,
    \label{eq:threesitefend}
\end{eqnarray}
 for gauge coupling strengths,
\begin{eqnarray}
  \mu \dfrac{d}{d\mu} \left(\dfrac{1}{g_0^2}\right)
  &=& \dfrac{1}{(4\pi)^2} 
      \left[ \dfrac{44}{3} - \dfrac{1}{6} \right],
      \label{eq:threesitegbegin}
  \\
  \mu \dfrac{d}{d\mu} \left(\dfrac{1}{g_1^2}\right)
  &=& \dfrac{1}{(4\pi)^2} 
      \left[ \dfrac{44}{3} - \dfrac{2}{6} \right],
  \\
  \mu \dfrac{d}{d\mu} \left(\dfrac{1}{g_2^2}\right)
  &=& \dfrac{1}{(4\pi)^2} 
      \left[ \phantom{\dfrac{44}{3}} - \dfrac{1}{6} \right],
      \label{eq:threesitegend}
\end{eqnarray}
 for ${\cal O}(p^4)$ terms,
\begin{eqnarray}
  \mu \dfrac{d}{d\mu} \alpha_{(i)1} 
  &=& \dfrac{1}{6(4\pi)^2}, 
  \label{eq:threesitealphabegin}
  \\
  \mu \dfrac{d}{d\mu} \alpha_{(i)2} 
  &=& \dfrac{1}{12(4\pi)^2}, 
  \\
  \mu \dfrac{d}{d\mu} \alpha_{(i)3} 
  &=& \dfrac{1}{12(4\pi)^2}, 
  \\
  \mu \dfrac{d}{d\mu} \alpha_{(i)4} 
  &=& -\dfrac{1}{6(4\pi)^2}, 
  \\
  \mu \dfrac{d}{d\mu} \alpha_{(i)5} 
  &=& -\dfrac{1}{12(4\pi)^2}, 
  \label{eq:threesitealphaend}
\end{eqnarray}
 and 
\begin{equation}
  \mu \dfrac{d}{d\mu} x_1 = \dfrac{3g_1^2}{(4\pi)^2} x_1,
  \label{eq:threesitedelocal}
\end{equation}
for the fermion delocalization operator.
Here we have assumed $\beta_{(2)} \ll 1$
and have ignored additional ${\cal O}(\beta_{(2)})$ terms in 
these RGEs.

Next we  solve the renormalization group equations assuming
\begin{equation}
  \beta_{(2)} \ll 1 .
\end{equation}
We find
\begin{eqnarray}
  f_1^2(\mu) &=& f_1^2(\Lambda)
  \exp\left[
    \int_\Lambda^\mu \dfrac{d\mu}{\mu} \dfrac{3}{(4\pi)^2} (g_0^2 + g_1^2 )
  \right],
  \\
  f_2^2(\mu) &=& f_2^2(\Lambda)
  \exp\left[
    \int_\Lambda^\mu \dfrac{d\mu}{\mu} \dfrac{3}{(4\pi)^2} (g_1^2
  +\frac{1}{2} g_2^2 )
  \right], 
  \\
  \beta_{(2)}(\mu) 
  & = & \dfrac{3}{4(4\pi)^2} \int_\Lambda^\mu \dfrac{d\mu}{\mu}
       g_2^2(\mu),
\label{eq:beta2rge}
\end{eqnarray}
for the ${\cal O}(p^2)$ terms,
\begin{eqnarray}
  \dfrac{1}{g_{0}^2(\mu)}
  &=& \dfrac{1}{g_{0}^2(\Lambda)}
     +\dfrac{87}{6(4\pi)^2} \ln \dfrac{\mu}{\Lambda}, 
  \\
  \dfrac{1}{g_{1}^2(\mu)}
  &=& \dfrac{1}{g_{1}^2(\Lambda)}
     +\dfrac{43}{3(4\pi)^2} \ln \dfrac{\mu}{\Lambda}, \label{eq:g1run235}
  \\
  \dfrac{1}{g_{2}^2(\mu)}
  &=& \dfrac{1}{g_{2}^2(\Lambda)}
     -\dfrac{1}{6(4\pi)^2} \ln \dfrac{\mu}{\Lambda}, 
\end{eqnarray}
for the gauge-coupling strengths, and
\begin{equation}
  x_1(\mu) = x_1(\Lambda) \exp \left[
    \int^\mu_\Lambda \dfrac{d\mu}{\mu} \dfrac{3g_1^2}{(4\pi)^2}
  \right], \label{eq:x1run237}
\end{equation}
for the delocalization parameter. We may similarly solve for the 
${\cal O}(p^4)$ coefficients. Here we list the results only for $\alpha_{(i)1}$'s
explicitly
\begin{eqnarray}
  \alpha_{(1)1}(\mu) 
    &=& \dfrac{1}{6(4\pi)^2} \ln \frac{\mu}{\Lambda}
       +\alpha_{(1)1}(\Lambda), \label{eq:a11run238}
    \\
  \alpha_{(2)1}(\mu) 
    &=& \dfrac{1}{6(4\pi)^2} \ln \frac{\mu}{\Lambda}
       +\alpha_{(2)1}(\Lambda). \label{eq:a21run239}
\end{eqnarray}
If we further assume
\begin{equation}
  g_0^2, g_2^2 \ll g_1^2, (4\pi)^2,
\label{eq:assume}
\end{equation}
we find
\begin{eqnarray}
  f_1^2(\mu) &=& f_1^2(\Lambda)
  \exp\left[
    \int_\Lambda^\mu \dfrac{d\mu}{\mu} \dfrac{3}{(4\pi)^2} g_1^2 
  \right],
  \\
  f_2^2(\mu) &=& f_2^2(\Lambda)
  \exp\left[
    \int_\Lambda^\mu \dfrac{d\mu}{\mu} \dfrac{3}{(4\pi)^2} g_1^2
  \right],
\end{eqnarray}
which, assuming $f^2_1(\Lambda)=f^2_2(\Lambda)$,  justifies the ansatz 
\begin{equation}
  f_1^2(\mu) = f_2^2(\mu),
\end{equation}
adopted in the discussion of the three site model in  Refs.\cite{SekharChivukula:2006cg} and \cite{Matsuzaki:2006wn}.
The assumption Eq.(\ref{eq:assume}) also makes it possible to neglect
the $\mu$ dependence of $g_2(\mu)$ in Eq.(\ref{eq:beta2rge}).  
We then find
\begin{equation}
  \beta_{(2)}(\mu) 
  = \dfrac{3}{4(4\pi)^2} g_2^2 \ln\frac{\mu}{\Lambda} .
  \label{eq:beta244}
\end{equation}

As demonstrated in the appendix, the three site RGE
equations for $\alpha_{(i)1-5}$ arise solely from Goldstone Boson loops, and are therefore identical
with those calculated \cite{Tanabashi:1993sr,Harada:2003jx} 
in hidden local symmetry \cite{Bando:1985ej,Bando:1985rf,Bando:1988ym,Bando:1988br,Harada:2003jx} models of QCD in the ``vector limit" \cite{Georgi:1989xy}.

\section{Matching to the Two Site Model and Running to Low Energies}

\label{sec:twosite}

In order to run to scales lower than $M_\rho$, we must integrate out the $\rho$-meson
and match the three site model to the two site electroweak chiral lagrangian which describes
the physics at scales below $M_\rho$. This
matching is most conveniently done in two steps: first we reformulate the effect of fermion delocalization
in terms of a redefinition of the chiral parameters in the three site lagrangian,
a procedure described in the following subsection, and then we explicitly  match the three site model
 to the two site model, as described in the second subsection. We conclude
 this section with a description of the running in the two site model for the energy range between
 $\mu=M_\rho$ and low energies $\mu=M_{H,ref}$.

\subsection{Field redefinitions and fermion delocalization}

We begin by reformulating the effect of fermion delocalization
in terms of a redefinition of the chiral parameters in the three site lagrangian with
brane-localized fermion couplings (to leading order in $x_1$).
The delocalized fermion coupling 
in Eq.(\ref{eq:fermion_coupling}) can be written in a localized manner 
\begin{equation}
  {\cal L}_f = -2J_W^{a\mu}
                \tr\left[T^a W_{\mu}\right]
               -2J_Y^{\mu} 
                \tr\left[T^3 B_{\mu}\right],
\end{equation}
if we redefine the gauge fields as
\begin{eqnarray}
  W_\mu &\equiv& V_{0\mu} + x_1 i (D_\mu U_1)U_1^\dagger, \\
  B_\mu &\equiv& V_{2\mu}.
\end{eqnarray}
Note that $W_\mu$ transforms under the gauge symmetry in a
manner similar to $V_{0\mu}$. 
Defining
\begin{equation}
  \tilde{D}_\mu U_1 
  \equiv \partial_\mu U_1 + i W_{\mu} U_1 - i U_1 V_{1\mu},
\end{equation}
we  find
\begin{equation}
  D_\mu U_1 = \dfrac{1}{1-x_1} \tilde{D}_\mu U_1 , 
\end{equation}
and obtain
\begin{equation}
  V_{0\mu} = W_{\mu} 
            - \dfrac{x_1}{1-x_1} i (\tilde{D}_\mu U_1) U_1^\dagger
  .
\end{equation}
Defining the $W$ field strength $W_{\mu\nu}$ as
\begin{equation}
  W_{\mu\nu} \equiv 
  \partial_\mu W_{\nu} - \partial_\nu W_{\mu} 
  + i [W_{\mu}, W_{\nu}]~,
\end{equation}
we find that $V_{0\mu\nu}$ can be expressed in terms of $W_\mu$,
\begin{equation}
  V_{0\mu\nu} = \dfrac{1}{1-x_1} W_{\mu\nu} 
               -\dfrac{x_1}{1-x_1} U_1 V_{1\mu\nu} U_1^\dagger
               -\dfrac{x_1}{(1-x_1)^2} 
                i[(\tilde{D}_\mu U_1)U_1^\dagger, 
                  (\tilde{D}_\nu U_1)U_1^\dagger ] 
               .
\end{equation}
The ${\cal O}(p^2)$ Lagrangian Eq.(\ref{eq:p2}) may then be written as
\begin{eqnarray}
  {\cal L}_2 &=& \dfrac{f_1^2}{4}(1+2x_1)\tr\left[
                   (\tilde{D}_\mu U_1)^\dagger  (\tilde{D}^\mu U_1)
                 \right] 
                +\dfrac{f_2^2}{4}\tr\left[
                   (D_\mu U_2)^\dagger  (D^\mu U_2)
                 \right]
             \nonumber\\
             & & -\dfrac{1}{2g_0^2}(1+2x_1)\tr\left[
                    W_{\mu\nu} W^{\mu\nu}
                  \right]
                 -\dfrac{1}{2g_1^2}\tr\left[
                    V_{1\mu\nu} V_1^{\mu\nu}
                  \right]
                 -\dfrac{1}{2g_2^2}\tr\left[
                    B_{\mu\nu} B^{\mu\nu}
                  \right]
             \nonumber\\
             & &
             +\dfrac{x_1}{g_0^2} \tr\left[
               W_{\mu\nu} U_1 V_1^{\mu\nu} U_1^\dagger
              \right]
             -2\dfrac{x_1}{g_0^2} i \tr\left[
               W_{\mu\nu} (\tilde{D}^\mu U_1)
               (\tilde{D}^\nu U_1)^\dagger \right]
             +{\cal O}(x_1^2)~.
\label{eq:lag2r}
\end{eqnarray}
The ${\cal O}(p^4)$ Lagrangians ${\cal L}_4^{(i)}$ become
\begin{eqnarray}
  {\cal L}_4^{(1)}
  &=& \alpha_{(1)1} \tr\left[ 
        W_{\mu\nu} U_1 V_{1}^{\mu\nu} U_1^\dagger
      \right]
  \nonumber\\
  & & -2i\alpha_{(1)2} \tr\left[
        (\tilde{D}_\mu U_1)^\dagger (\tilde{D}_\nu U_1) V_{1}^{\mu\nu}
      \right]
  \nonumber\\
  & & -2i\alpha_{(1)3} \tr\left[
        W^{\mu\nu} (\tilde{D}_\mu U_1) (\tilde{D}_\nu U_1)^\dagger
      \right] 
  \nonumber\\
  & & +\alpha_{(1)4} 
       \tr\left[(\tilde{D}_\mu U_1) (\tilde{D}_\nu U_1)^\dagger \right] 
       \tr\left[(\tilde{D}^\mu U_1) (\tilde{D}^\nu U_1)^\dagger \right] 
  \nonumber\\
  & & +\alpha_{(1)5} 
       \tr\left[(\tilde{D}_\mu U_1) (\tilde{D}^\mu U_1)^\dagger \right] 
       \tr\left[(\tilde{D}_\nu U_1) (\tilde{D}^\nu U_1)^\dagger
  \right]
     + {\cal O}(x_1 \alpha_{(i)j}),
\label{eq:lag41r}
\end{eqnarray}
and
\begin{eqnarray}
  {\cal L}_{4}^{(2)} 
  &=& \alpha_{(2)1} \tr\left[ 
        V_{1\mu\nu} U_2 B^{\mu\nu} U_2^\dagger
      \right]
  \nonumber\\
  & & -2i\alpha_{(2)2} \tr\left[
        (D_\mu U_2)^\dagger (D_\nu U_2) B^{\mu\nu}
      \right]
  \nonumber\\
  & & -2i\alpha_{(2)3} \tr\left[
        V_{1}^{\mu\nu} (D_\mu U_2) (D_\nu U_2)^\dagger
      \right] 
  \nonumber\\
  & & +\alpha_{(2)4} 
       \tr\left[(D_\mu U_2) (D_\nu U_2)^\dagger \right] 
       \tr\left[(D^\mu U_2) (D^\nu U_2)^\dagger \right] 
  \nonumber\\
  & & +\alpha_{(2)5} 
       \tr\left[(D_\mu U_2) (D^\mu U_2)^\dagger \right] 
       \tr\left[(D_\nu U_2) (D^\nu U_2)^\dagger \right]
      + {\cal O}(x_1 \alpha_{(i)j}).
\label{eq:lag42r2}
\end{eqnarray}
We note that the Lagrangian Eq.(\ref{eq:lag2r}) now contains 
${\cal O}(p^4)$ terms.
We thus rearrange Eq.(\ref{eq:lag2r}) and Eq.(\ref{eq:lag41r})
as
\begin{eqnarray}
  \tilde{{\cal L}}_2 &=& \dfrac{\tilde{f}_1^2}{4}\tr\left[
                   (\tilde{D}_\mu U_1)^\dagger  (\tilde{D}^\mu U_1)
                 \right] 
                +\dfrac{f_2^2}{4}\tr\left[
                   (D_\mu U_2)^\dagger  (D^\mu U_2)
                 \right]
             \nonumber\\
             & & -\dfrac{1}{2\tilde{g}_0^2}\tr\left[
                    W_{\mu\nu} W^{\mu\nu}
                  \right]
                 -\dfrac{1}{2g_1^2}\tr\left[
                    V_{1\mu\nu} V_1^{\mu\nu}
                  \right]
                 -\dfrac{1}{2g_2^2}\tr\left[
                    B_{\mu\nu} B^{\mu\nu}
                  \right],
\label{eq:lag2r2}
\end{eqnarray}
and
\begin{eqnarray}
  \tilde{{\cal L}}_4^{(1)}
  &=& \tilde{\alpha}_{(1)1} \tr\left[ 
        W_{\mu\nu} U_1 V_{1}^{\mu\nu} U_1^\dagger
      \right]
  \nonumber\\
  & & -2i\alpha_{(1)2} \tr\left[
        (\tilde{D}_\mu U_1)^\dagger (\tilde{D}_\nu U_1) V_{1}^{\mu\nu}
      \right]
  \nonumber\\
  & & -2i\tilde{\alpha}_{(1)3} \tr\left[
        W^{\mu\nu} (\tilde{D}_\mu U_1) (\tilde{D}_\nu U_1)^\dagger
      \right] 
  \nonumber\\
  & & +\alpha_{(1)4} 
       \tr\left[(\tilde{D}_\mu U_1) (\tilde{D}_\nu U_1)^\dagger \right] 
       \tr\left[(\tilde{D}^\mu U_1) (\tilde{D}^\nu U_1)^\dagger \right] 
  \nonumber\\
  & & +\alpha_{(1)5} 
       \tr\left[(\tilde{D}_\mu U_1) (\tilde{D}^\mu U_1)^\dagger \right] 
       \tr\left[(\tilde{D}_\nu U_1) (\tilde{D}^\nu U_1)^\dagger
  \right]
     + {\cal O}(x_1 \alpha_{(i)j}),
\label{eq:lag41r2}
\end{eqnarray}
with
\begin{eqnarray}
  \tilde{f}_1^2 &=& f_1^2(1+2x_1), \\
  \dfrac{1}{\tilde{g}_0^2} &=& \dfrac{1+2x_1}{g_0^2} , \\
  \tilde{\alpha}_{(1)1} &=& \alpha_{(1)1} +\dfrac{x_1}{g_0^2}, \\
  \tilde{\alpha}_{(1)3} &=& \alpha_{(1)3} +\dfrac{x_1}{g_0^2}.
\end{eqnarray}

\subsection{Matching with the two site model}

\EPSFIGURE[tb]{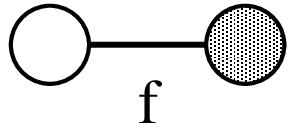,width=4cm}
 {Moose diagram \protect\cite{Georgi:1985hf}  for the two site electroweak chiral Lagrangian
 \protect\cite{Appelquist:1980ae,Appelquist:1980vg,Longhitano:1980iz,Longhitano:1980tm,Appelquist:1993ka}, an
  $SU(2)\times U(1)$ gauged nonlinear sigma model. The $SU(2)_W$ gauge group is shown
  as an open circle; the $U(1)_Y$ gauge group as a shaded circle. The link represents the nonlinear
  sigma model field $U$, with $f$-constant $f$.
 \label{fig:twosite2ew}}

Below the KK mass scale, phenomenology of the three site
model can be described by a two site model, i.e., the electroweak chiral
Lagrangian illustrated in Figure~\ref{fig:twosite2ew}, with the ${\cal O}(p^2)$ Lagrangian
\begin{equation}
  {\cal L}_2 = 
    \dfrac{f^2}{4} 
    \tr\left[(D_\mu U)^\dagger (D^\mu U) \right]
   -\dfrac{1}{2g_W^2}
    \tr\left[W_{\mu\nu}W^{\mu\nu} \right]
   -\dfrac{1}{2g_Y^2}
    \tr\left[B_{\mu\nu}B^{\mu\nu} \right]~,
\end{equation}
where
\begin{equation}
  D_\mu U = \partial_\mu U + i W_{\mu} U
             - i U B_{\mu}, \qquad
  U \equiv U_1 U_2 .
\end{equation}
There also exists a custodial symmetry violating ${\cal O}(p^2)$
operator 
\begin{equation}
  {\cal L}_2' = \beta \dfrac{f^2}{4} 
    \tr\left[ U^\dagger (D_\mu U)\tau^3 \right]
    \tr\left[ U^\dagger (D^\mu U)\tau^3 \right]~,
\end{equation}
and the ${\cal O}(p^4)$ operators
\begin{eqnarray}
  {\cal L}_{4}
  &=& \alpha_{1} \tr\left[ 
        W_{\mu\nu} U B^{\mu\nu} U^\dagger
      \right]
  \nonumber\\
  & & -2i\alpha_{2} \tr\left[
        (D_\mu U)^\dagger (D_\nu U) B_{\mu\nu}
      \right]
  \nonumber\\
  & & -2i\alpha_{3} \tr\left[
        W^{\mu\nu} (D_\mu U) (D_\nu U)^\dagger
      \right] 
  \nonumber\\
  & & +\alpha_{4} 
       \tr\left[(D_\mu U) (D_\nu U)^\dagger \right] 
       \tr\left[(D^\mu U) (D^\nu U)^\dagger \right] 
  \nonumber\\
  & & +\alpha_{5} 
       \tr\left[(D_\mu U) (D^\mu U)^\dagger \right] 
       \tr\left[(D_\nu U) (D^\nu U)^\dagger \right] ~.
\end{eqnarray}

We next perform matching between three site and two site models.
We will assume
\begin{equation}
  \beta_{(2)}, \alpha_{(i)1,2,3,4,5}, x_1 \ll 1, 
\end{equation}
and treat these parameters in a perturbative manner. In the limit $g_0, g_2 \ll g_1$,
the mass-eigenstate $\rho$ field being integrated out is approximately the same
as the field  $V_{1\mu}$ of the three site model. Using the equations of motion arising
from Eq.~(\ref{eq:lag2r2}), to leading order in the derivative expansion
\begin{equation}
\frac{\delta \tilde{\cal L}_2}{\delta V_{1\mu}} = \partial_\nu \frac{\delta \tilde{\cal L}_2}
{\delta(\partial_\nu V_{1\mu})}\approx 0~,
\end{equation}
this field may be expressed as
\begin{equation}
  V_{1\mu} = \dfrac{1}{\tilde{f}_1^2+f_2^2}\left[
    \tilde{f}_1^2 U_1^\dagger W_{\mu} U_1 + f_2^2 U_2 B_{\mu} U_2^\dagger
   -i\tilde{f}_1^2 U_1^\dagger \partial_\mu U_1 
   -if_2^2 U_2 \partial_\mu U_2^\dagger \right].
\end{equation}

From this we obtain
\begin{eqnarray}
  U_1^\dagger \tilde{D}_\mu U_1 
  &=& \dfrac{f_2^2}{\tilde{f}_1^2+f_2^2} \left[
         U_1^\dagger \partial_\mu U_1 - U_2 \partial_\mu U_2^\dagger
      +i U_1^\dagger W_{\mu} U_1 - i U_2 B_\mu U_2^\dagger
      \right] 
  \nonumber\\
  &=& \dfrac{f_2^2}{\tilde{f}_1^2+f_2^2} 
      U_1^\dagger (D_\mu U) U_2^\dagger,
  \\
  U_2 (D_\mu U_2)^\dagger
  &=& \dfrac{\tilde{f}_1^2}{\tilde{f}_1^2+f_2^2}\left[
        -U_1^\dagger \partial_\mu U_1 + U_2\partial_\mu U_2^\dagger
        -iU_1^\dagger W_\mu U_1 + i U_2 B_\mu U_2^\dagger
      \right] 
  \nonumber\\
  &=& -\dfrac{\tilde{f}_1^2}{\tilde{f}_1^2+f_2^2}
      U_1^\dagger (D_\mu U) U_2^\dagger,
  \\
  V_{1\mu\nu} 
  &=& \dfrac{\tilde{f}_1^2}{\tilde{f}_1^2+f_2^2} 
      U_1^\dagger W_{\mu\nu} U_1
     +\dfrac{f_2^2}{\tilde{f}_1^2+f_2^2} 
      U_2 B_{\mu\nu} U_2^\dagger
  \nonumber\\
  & & +\dfrac{\tilde{f}_1^2 f_2^2}{\left(\tilde{f}_1^2+f_2^2\right)^2} 
      i U_2 \left[ U^\dagger (D_\mu U), U^\dagger (D_\nu U) \right] 
        U_2^\dagger .
\end{eqnarray}
Putting these into Eqs. (\ref{eq:beta2}), (\ref{eq:lag42r2}),
(\ref{eq:lag2r2}) and (\ref{eq:lag41r2}), we match the 
two site model to the three site model, and find matching conditions
\begin{eqnarray}
  \dfrac{1}{f^2} &=& \dfrac{1}{\tilde{f}_1^2} + \dfrac{1}{f_2^2}
  \simeq \dfrac{1}{f_1^2} + \dfrac{1}{f_2^2} ,
  \\
  \beta &=& \beta_{(2)} \dfrac{\tilde{f}_1^2}{\tilde{f}_1^2 + f_2^2}
  \simeq \beta_{(2)} \dfrac{f_1^2}{f_1^2 + f_2^2} \label{eq:beta:329}
  , \\
  \dfrac{1}{g_W^2} 
  &=& \dfrac{1}{\tilde{g}_0^2} + \dfrac{1}{g_1^2} 
      \left(\dfrac{\tilde{f}_1^2}{\tilde{f}_1^2+f_2^2}\right)^2 
     -2\tilde{\alpha}_{(1)1} 
      \left(\dfrac{\tilde{f}_1^2}{\tilde{f}_1^2+f_2^2}\right)
  \nonumber\\
  &\simeq& \dfrac{1}{g_0^2}, 
  \\
  \dfrac{1}{g_Y^2} 
  &=& \dfrac{1}{g_2^2} + \dfrac{1}{g_1^2} 
      \left(\dfrac{f_2^2}{\tilde{f}_1^2+f_2^2}\right)^2 
     -2\alpha_{(2)1} \left(\dfrac{f_2^2}{\tilde{f}_1^2+f_2^2}\right)
  \nonumber\\
  &\simeq& \dfrac{1}{g_2^2},
  \end{eqnarray}
\begin{eqnarray}
  \alpha_{1}
  &=& -\dfrac{1}{g_1^2} 
       \left(\dfrac{\tilde{f}_1 f_2}{\tilde{f}_1^2+f_2^2}\right)^2
      +\tilde{\alpha}_{(1)1} \left(\dfrac{f_2^2}{\tilde{f}_1^2+f_2^2}\right)
      +\alpha_{(2)1}
       \left(\dfrac{\tilde{f}_1^2}{\tilde{f}_1^2+f_2^2}\right) 
  \nonumber\\
  &\simeq& -\dfrac{1}{g_1^2} 
       \dfrac{f_1^2 f_2^2}{(f_1^2+f_2^2)^2}
      +\left(\alpha_{(1)1}+\dfrac{x_1}{g_0^2}\right)
       \dfrac{f_2^2}{f_1^2+f_2^2}
      +\alpha_{(2)1}
       \dfrac{f_1^2}{f_1^2+f_2^2}
  ,
\label{eq:alpha1m}
  \\
  \alpha_2 
  &\simeq& 
      -\dfrac{1}{g_1^2}\dfrac{f_1^2 f_2^4}{(f_1^2+f_2^2)^3}
      +\alpha_{(2)1} \dfrac{f_1^2 f_2^2}{(f_1^2+f_2^2)^2}
  \nonumber\\
  & &
      +\alpha_{(1)2} \dfrac{f_2^6}{(f_1^2+f_2^2)^3}
      +\alpha_{(2)2} \dfrac{f_1^4}{(f_1^2+f_2^2)^2}
      +\alpha_{(2)3} \dfrac{f_1^4 f_2^2}{(f_1^2+f_2^2)^3},
\label{eq:alpha2m}
  \\ 
  \alpha_3 
  &\simeq& 
      -\dfrac{1}{g_1^2}\dfrac{f_1^4 f_2^2}{(f_1^2+f_2^2)^3}
      +\left(\alpha_{(1)1}+\dfrac{x_1}{g_0^2}\right)
       \dfrac{f_1^2 f_2^2}{(f_1^2+f_2^2)^2}
  \nonumber\\
  & &
      +\alpha_{(1)2} \dfrac{f_1^2 f_2^4}{(f_1^2+f_2^2)^3}
      +\left(\alpha_{(1)3} +\dfrac{x_1}{g_0^2}\right)
       \dfrac{f_2^4}{(f_1^2+f_2^2)^2}
      +\alpha_{(2)3} \dfrac{f_1^6}{(f_1^2+f_2^2)^3},
\label{eq:alpha3m}
\\
  \alpha_4 
  &\simeq & 
     \frac{1}{g^2_1} \Bigg( \frac{f_1f_2}{f_1^2+f_2^2} \Bigg)^4 
   - 2 \alpha_{(1)2} \frac{f_1^2f_2^6}{(f_1^2+f_2^2)^4} 
   - 2 \alpha_{(2)3} \frac{f_1^6f_2^2}{(f_1^2+f_2^2)^4} 
   \nonumber \\ 
  & & 
     + \alpha_{(1)4} \Bigg(\frac{f_2^2}{f_1^2+f_2^2}\Bigg)^4
    + \alpha_{(2)4} \Bigg( \frac{f_1^2}{f_1^2+f_2^2} \Bigg)^4 
    \,, \\ 
  \alpha_5 
  &\simeq & 
    - \frac{1}{g^2_1} \Bigg( \frac{f_1f_2}{f_1^2+f_2^2} \Bigg)^4 
   + 2 \alpha_{(1)2} \frac{f_1^2f_2^6}{(f_1^2+f_2^2)^4} 
    + 2 \alpha_{(2)3} \frac{f_1^6f_2^2}{(f_1^2+f_2^2)^4} 
   \nonumber \\ 
  & & 
    + \alpha_{(1)5} \Bigg(\frac{f_2^2}{f_1^2+f_2^2}\Bigg)^4
    + \alpha_{(2)5} \Bigg( \frac{f_1^2}{f_1^2+f_2^2} \Bigg)^4~,
    \label{eq:newlabel}
\end{eqnarray}
where we match at a scale
\begin{equation}
  \mu = M_{\rho} \simeq g_1 \dfrac{\sqrt{f_1^2+f_2^2}}{2} .
\label{eq:rhomassmu}
\end{equation}

Finally, we note that we can make contact with prior results by setting $f_1=f_2$ and $\alpha_{(i)j}=0$.
In this limit, Eqs.(\ref{eq:alpha1m})--(\ref{eq:newlabel}) lead to
\begin{eqnarray}
  \alpha_1 = -\dfrac{1}{4g_1^2} &+& \dfrac{x_1}{2g_0^2}, \qquad 
  \alpha_2 = -\dfrac{1}{8g_1^2}, \qquad 
  \alpha_3 = -\dfrac{1}{8g_1^2} + \dfrac{x_1}{2g_0^2}, \\
\alpha_4 & = &  \frac{1}{16g_1^2}, \qquad \qquad 
\alpha_5  =   - \frac{1}{16g_1^2} ,
\end{eqnarray}
which are consistent with Table~2 given in ~\cite{SekharChivukula:2006cg}. 

\subsection{Solutions of RGE in the two site model}

\label{sec:twositescaling}

We next consider the renormalization group flow in the two site model, from
a scale $\mu=M_\rho$ to low-energy, $\mu=M_{H,{\rm ref}}$.
The renormalization group equations of the two site model, the derivation
of which is described in \ref{sec:21model},  are given by
\begin{eqnarray}
  \mu \dfrac{d}{d\mu} f^2 
    &=& \dfrac{3}{(4\pi)^2} (g_W^2 + \frac{1}{2} g_Y^2) f^2, 
    \label{eq:twositebegin}
  \\
  \mu \dfrac{d}{d\mu} \left( \beta f^2 \right) 
    &=& \dfrac{3}{4(4\pi)^2} g_Y^2 f^2,
\end{eqnarray}
\begin{eqnarray}
  \mu \dfrac{d}{d\mu} \left(\dfrac{1}{g_W^2}\right)
  &=& \dfrac{1}{(4\pi)^2} 
      \left[ \dfrac{44}{3} - \dfrac{1}{6} \right],
  \\
  \mu \dfrac{d}{d\mu} \left(\dfrac{1}{g_Y^2}\right)
  &=& \dfrac{1}{(4\pi)^2} 
      \left[ \phantom{\dfrac{44}{3}} - \dfrac{1}{6} \right],
\end{eqnarray}
\begin{eqnarray}
  \mu \dfrac{d}{d\mu} \alpha_{1} 
  &=& \dfrac{1}{6(4\pi)^2}, 
  \\
  \mu \dfrac{d}{d\mu} \alpha_{2} 
  &=& \dfrac{1}{12(4\pi)^2}, 
  \\
  \mu \dfrac{d}{d\mu} \alpha_{3} 
  &=& \dfrac{1}{12(4\pi)^2}, 
  \\
  \mu \dfrac{d}{d\mu} \alpha_{4} 
  &=& -\dfrac{1}{6(4\pi)^2}, 
  \\
  \mu \dfrac{d}{d\mu} \alpha_{5} 
  &=& -\dfrac{1}{12(4\pi)^2}. 
  \label{eq:twositeend}
\end{eqnarray}

We solve these equations assuming
\begin{equation}
  \beta \ll 1 .
\end{equation}
We find\footnote{This result is identical to that found \protect\cite{Longhitano:1980iz},
in the case of the effective low-energy theory for a standard model with a heavy Higgs
boson.}
\begin{equation}
  \beta(\mu) 
  = \frac{3}{4}\dfrac{g_Y^2}{(4\pi)^2} \ln \dfrac{\mu}{M_\rho}
   +\beta(M_\rho),
   \label{eq:beta353}
\end{equation}
and
\begin{eqnarray}
  \alpha_{1}(\mu) 
    &=& \dfrac{1}{6(4\pi)^2} \ln \frac{\mu}{M_\rho}
       +\alpha_{1}(M_\rho), \label{eq:alpha354}
 \end{eqnarray}
and similarly for the other $\alpha_i$.

\section{$\alpha S$ and $\alpha T$}

\label{sec:st}

Using the results of Sections 2 and 3, we may compute the
values of $\alpha S$ and $\alpha T$, which are defined as
\begin{eqnarray}
  \alpha S &=& -16\pi\alpha\, \alpha_1(\mu=M_{H,\rm ref}), \label{eq:sdeff}\\
  \alpha T &=& 2\beta(\mu=M_{H,\rm ref}). \label{eq:tdeff}
\end{eqnarray}

We begin with Eqn. (\ref{eq:sdeff}) and use the RGE equations to evaluate $\alpha  S$ at successively higher energy scales, and eventually
compare it with the results in Ref.  \cite{Matsuzaki:2006wn}.   
First, we can use Eqn. (\ref{eq:alpha354})
to run up from $M_{H,\rm ref}$ to $M_\rho$
\begin{equation}
 \alpha S = -16\pi\alpha\, \alpha_1(M_\rho) 
        - \dfrac{\alpha}{6\pi} \ln \dfrac{M_{H,\rm ref}}{M_\rho}.
\end{equation}
Taking $f_1 = f_2$, we apply Eqn. (\ref{eq:alpha1m}) to match from the two-site to the three-site regime
\begin{equation}
\alpha  S \simeq \dfrac{4\pi\alpha}{g_1^2(M_\rho)}
        -8\pi \alpha\,\alpha_{(1)1}(M_\rho) 
        -8\pi \alpha\,\alpha_{(2)1}(M_\rho) 
        - \dfrac{8\pi \alpha x_1(M_\rho)}{g_0^2(M_\rho)}
        - \dfrac{\alpha}{6\pi} \ln \dfrac{M_{H,\rm ref}}{M_\rho}.
\end{equation}
We employ Eqns. (\ref{eq:x1run237}, \ref{eq:a11run238}, \ref{eq:a21run239}) to run $x_1$, $\alpha_{(1)1}$, and $\alpha_{(2)1}$ up to scale $\Lambda$
\begin{eqnarray}
\alpha  S    &\simeq& \dfrac{4\pi\alpha}{g_1^2(M_\rho)}
        - \dfrac{\alpha}{6\pi} \ln \dfrac{M_{H,\rm ref}}{M_\rho}
        - \dfrac{\alpha}{6\pi} \ln \dfrac{M_\rho}{\Lambda}
        -8\pi \alpha\,\alpha_{(1)1}(\Lambda) 
        -8\pi \alpha\,\alpha_{(2)1}(\Lambda) 
    \nonumber\\
    & & -\dfrac{8\pi\alpha  x_1(\Lambda)}{g_0^2(\Lambda)} \exp\left[
         \int^{M_\rho}_\Lambda \dfrac{d\mu}{\mu} 
         \dfrac{3g_1^2(\mu)}{(4\pi)^2}
         \right],
\end{eqnarray}
and then employ Eqn. (\ref{eq:g1run235}) to run $g_1$ up to scale $\Lambda$
\begin{eqnarray}
\alpha  S    &\simeq& \dfrac{4\pi\alpha}{g_1^2(\Lambda)}
        - \dfrac{8\pi \alpha x_1(\Lambda)}{g_0^2(\Lambda)}
        - \dfrac{\alpha}{6\pi} \ln \dfrac{M_{H,\rm ref}}{M_\rho}
        + \dfrac{41\alpha }{12\pi} \ln \dfrac{M_\rho}{\Lambda}
        - \dfrac{3\alpha}{2\pi} \dfrac{x_1 g_1^2}{g_0^2}
          \ln \dfrac{M_\rho}{\Lambda}
   \nonumber\\
   & & 
        -8\pi \alpha\, \alpha_{(1)1}(\Lambda) 
        -8\pi \alpha\,\alpha_{(2)1}(\Lambda).
\label{eq:Sresult}
\end{eqnarray}
In the first line of Eq.(\ref{eq:Sresult}), we have used the expansion
\begin{equation}
  \exp\left[
         \int^{M_\rho}_\Lambda \dfrac{d\mu}{\mu} 
         \dfrac{3g_1^2(\mu)}{(4\pi)^2}
  \right]
  \simeq 1 + \dfrac{3}{(4\pi)^2} g_1^2 \ln \dfrac{M_\rho}{\Lambda}.
\end{equation}
In order to make connection with the results of ref. \cite{Matsuzaki:2006wn}, we note that 
at tree-level
\begin{equation}
M^2_W \approx\frac{g^2_0 f^2}{4} = \frac{g^2_0 f^2_1}{8}~,
\end{equation}
and, therefore  ({\it c.f.} Eq.~(\ref{eq:rhomassmu})),
\begin{equation}
\frac{g^2_1}{g^2_0} = \frac{M^2_\rho}{4 M^2_W}~,
\end{equation}
so that the first two terms of Eqn. (\ref{eq:Sresult}) may be rewritten as
\begin{equation}
\dfrac{4\pi\alpha}{g_1^2(\Lambda)}
        - \dfrac{8\pi \alpha x_1(\Lambda)}{g_0^2(\Lambda)}
        = \left[\frac{4 s^2 M^2_W}{M^2_\rho}
\left(1- \frac{x_1 M^2_\rho}{2 M^2_W}\right)\right]_{\mu=\Lambda} ~,
\end{equation}
to this order.

Similarly, starting from Eqn. (\ref{eq:tdeff}), we can use the RGE equations from the
previous sections to evaluate $\alpha  T$ at higher energy scales, and eventually
compare it with the results in Ref.  \cite{Matsuzaki:2006wn}.   First, we can use Eqn. (\ref{eq:beta353})
to run up from $M_{H,\rm ref}$ to $M_\rho$
\begin{equation}
 \alpha T = 2\beta(M_\rho) 
       +\dfrac{3 g_Y^2}{2(4\pi)^2} \ln \dfrac{M_{H,\rm ref}}{M_\rho}.
\end{equation}
Taking $f_1 = f_2$, we apply Eqn. (\ref{eq:beta:329}) to match from the two-site to the three-site regime
\begin{equation}
 \alpha T = \beta_{(2)}(M_\rho)
       +\dfrac{3 g_Y^2}{2(4\pi)^2} \ln \dfrac{M_{H,\rm ref}}{M_\rho},
\end{equation}
and then employ Eqn. (\ref{eq:beta244}) to run up to scale $\Lambda$
\begin{eqnarray}
  \alpha T 
   &=& \beta_{(2)}(\Lambda)
       +\dfrac{3 g_2^2}{4(4\pi)^2} \ln \dfrac{M_\rho}{\Lambda}
       +\dfrac{3 g_Y^2}{2(4\pi)^2} \ln \dfrac{M_{H,\rm ref}}{M_\rho}
    \nonumber\\
    &\simeq&
        \beta_{(2)}(\Lambda) + \dfrac{3 g_Y^2}{4(4\pi)^2} \ln \dfrac{M_\rho}{\Lambda}
       +\dfrac{3 g_Y^2}{2(4\pi)^2} \ln \dfrac{M_{H,\rm ref}}{M_\rho}.
\label{eq:Tresult}
\end{eqnarray}

We can now see that our leading-log
expressions for $\alpha S$ (\ref{eq:Sresult}) 
and $\alpha T$ (\ref{eq:Tresult}) correspond exactly to those found in \cite{Matsuzaki:2006wn}.
Note that the calculations in \cite{Matsuzaki:2006wn} are performed in
't Hooft-Feynman gauge, whereas those reported here are performed in Landau
gauge. The correspondence of these calculations is therefore an explicit demonstration
of the gauge-invariance of the results.  In addition, as noted in \cite{Matsuzaki:2006wn}, the $M_{H,{\rm ref}}$ dependence of the results in Eqns.~(\ref{eq:Sresult}) and (\ref{eq:Tresult})  matches that of
the low-energy effective theory of the standard model with a heavy Higgs
boson \cite{Appelquist:1980ae,Appelquist:1980vg,Longhitano:1980iz,Longhitano:1980tm},
and therefore these contributions are model-independent and match the dependence on the
reference Higgs-boson mass present in the experimental determinations of
$\alpha S$ and $\alpha T$.
 
Our expressions for the $S$ and $T$ parameters  include (a) the tree-level contributions, which are of order $N$, (b) the chiral corrections which are order $1$ but enhanced by chiral logs, and (c) the effects of the $p^4$ counterterms, which are simply order $1$.  While the coefficients of the counterterms are unknown, it is important to note that their magnitude is sub-leading in the large-$N$ expansion.  
A detailed investigation of the size of the chiral log corrections, and of
the corresponding phenomenological bounds on the three site model,
is underway \cite{unpublished}.

\section{Summary}

In this paper we have computed  the one-loop chiral logarithmic corrections to
all ${\cal O}(p^4)$ counterterms in the three site Higgsless model.
The calculation is performed using the background
field method for both the chiral- and gauge-fields, and using Landau gauge
for the quantum fluctuations of the gauge fields. For the chiral parameters
$\alpha_{(i)1-5}$, the contributions to the RGE equations arise solely from 
Goldstone Boson loops, and are therefore identical
with those calculated \cite{Tanabashi:1993sr} 
in hidden local symmetry \cite{Bando:1985ej,Bando:1985rf,Bando:1988ym,Bando:1988br,Harada:2003jx} models of QCD in the ``vector limit" \cite{Georgi:1989xy}.
Our results agree with previous
calculations \cite{Matsuzaki:2006wn} of the chiral-logarithmic corrections to the $S$ and $T$
parameters  in 't Hooft-Feynman gauge.

\bigskip

\centerline{\bf Acknowledgments}

We thank Kaoru Hagiwara and Qi-Shu Yan for
discussions on the renormalization of $SU(2)\times U(1)$ gauged
nonlinear sigma model. 
 The visit of S.M. to Michigan State University which made
this collaboration possible was fully supported by
the fund of The Mitsubishi Foundation through Koichi Yamawaki, and
this work received the preliminary report number DPNU-06-12.
R.S.C. and E.H.S. are supported in part by the US National Science Foundation under
grant  PHY-0354226.  M.T.'s work is supported in part by the JSPS Grant-in-Aid for
Scientific Research No.16540226

\appendix 
\renewcommand\theequation{\Alph{section}.\arabic{equation}}

\section{RGE Calculations}

\label{sec:rgeappendix}

In this appendix, we outline the calculation of the renormalization group
equations used in this paper. All calculations are performed using dimensional
regularization, renormalized using modified minimal subtraction ($\overline{\rm MS}$).
In order to facilitate the identification of the counterterms required, the calculations 
are performed using the background field method for both the chiral- and gauge-fields.
Landau gauge, in which all unphysical Goldstone bosons and ghost fields are massless,
is used for the quantum fluctuations of the gauge fields.

The calculation is carried out in three steps. In the next subsection, \ref{sec:22model},  we discuss
the RGE calculation in the context of a gauged $SU(2) \times SU(2)$ model. In the following subsection,
\ref{sec:21model}, we discuss the RGE calculation in a gauged $SU(2) \times U(1)$ model. Finally,
in the last subsection, \ref{sec:threesiteappendix}, we show how these results may be combined
to obtain the RGE equations for the three site gauged $SU(2) \times SU(2) \times U(1)$ model.
The results of the last subsection yield the equations necessary for section
\ref{sec:threesiterunning}: running from high-energies,
$\mu = \Lambda$, to intermediate energies, $\mu=M_\rho$. The results of the second
subsection directly yield the equations necessary for section \ref{sec:twositescaling}: running
between intermediate energies and low energies, $\mu=M_{H,{\rm ref}}$.

The divergences of the scalar one-loop integrals which appear are discussed, in general
form, in appendix \ref{sec:divergences}.

\subsection{$SU(2)\times SU(2)$ gauged nonlinear sigma model}
\label{sec:22model}

The one-loop renormalization of the operators \cite{Appelquist:1980ae,Appelquist:1980vg,Longhitano:1980iz,Longhitano:1980tm,Appelquist:1993ka} in the $SU(2)\times SU(2) / SU(2)$ chiral Lagrangian was first undertaken in Refs.  \cite{Weinberg:1978kz,Gasser:1983yg,Gasser:1984gg} and later discussed in the context of electroweak physics \cite{Herrero:1993nc,Dobado:1997jx}.  Here we review the  the RGE calculation in a gauged $SU(2) \times SU(2)$ model using the background field method for
the chiral- and gauge-fields,  and add a calculation of the renormalization of the operator associated with fermion delocalization effects. 

\EPSFIGURE[htbp]
{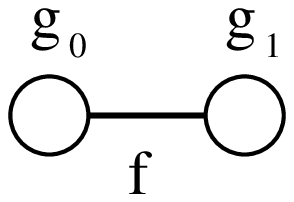,width=3cm}
{Moose diagram for the $SU(2)\times SU(2)$ gauged nonlinear
  sigma model analyzed in this subsection. $SU(2)$ gauge groups are shown
  as open circles.  \label{fig:twosite1}
  }

We begin by considering an $SU(2)\times SU(2)$ gauged nonlinear sigma
model shown in a moose diagram Figure~\ref{fig:twosite1}.
The lowest order (${\cal O}(p^2)$) Lagrangian of this model is given
by
\begin{equation}
  {\cal L}_2 = 
    \dfrac{f^2}{4} 
    \tr\left[(D_\mu U)^\dagger (D^\mu U) \right]
   -\dfrac{1}{2g_0^2}
    \tr\left[V_{0\mu\nu}V_0^{\mu\nu} \right]
   -\dfrac{1}{2g_1^2}
    \tr\left[V_{1\mu\nu}V_1^{\mu\nu} \right],
\label{eq:22p2}
\end{equation}
with $U$ being a chiral field,
\begin{equation}
  U \equiv \exp \left[ \dfrac{2i \pi^a T^a}{f} \right], \qquad
  T^a \equiv \dfrac{\tau^a}{2}. 
\end{equation}
The gauge field strengths $V_{0\mu\nu}$ and $V_{1\mu\nu}$ are defined
as
\begin{equation}
  V_{0\mu\nu} \equiv \partial_\mu V_{0\nu} - \partial_\nu V_{0\mu} 
                    +i[ V_{0\mu}, V_{0\nu}], \qquad
  V_{1\mu\nu} \equiv \partial_\mu V_{1\nu} - \partial_\nu V_{1\mu} 
                    +i[ V_{1\mu}, V_{1\nu}],
\end{equation}
with $V_{0\mu}$ and $V_{1\mu}$ being the gauge fields at site-0 and
site-1 in the moose diagram,
\begin{equation}
  V_{0\mu} \equiv V^a_{0\mu} T^a, \qquad
  V_{1\mu} \equiv V^a_{1\mu} T^a.
\end{equation}
The covariant derivative $D_\mu U$ is given by
\begin{equation}
  D_\mu U \equiv \partial_\mu U + i V_{0\mu} U - i U V_{1\mu}.
\end{equation}

We calculate one-loop divergences using the background field
formalism in order to maintain  manifest chiral and gauge invariance.
For this purpose we decompose the chiral field $U$ into a 
background field $\bar{U}$ and a fluctuating quantum Nambu-Goldstone Boson
(NGB) field $u^a$,
\begin{equation}
  U = \bar{U} \exp\left[ \dfrac{ 2i u^a T^a}{f} \right].
\end{equation}
In a similar manner, we decompose the site-1 gauge field $V_{1\mu}$
into a background field $\bar{V}_{1\mu}$ and a quantum field $v_{1\mu}$,
\begin{equation}
  V_{1\mu} = \bar{V}_{1\mu} + g_1 v_{1\mu}.
\end{equation}
For the site-0 gauge field $V_{0\mu}$, it is convenient to first make a
gauge transformation,
\begin{equation}
  V_{0\mu} \to 
  V'_{0\mu} = \bar{U}^\dagger V_{0\mu} \bar{U}
   - i \bar{U}^\dagger \partial_\mu \bar{U},
   \label{eq:A8}
\end{equation}
and then decompose it into a background field $\bar{V}_{0\mu}$ and
a quantum field $v_{0\mu}$,
\begin{equation}
  V'_{0\mu} = \bar{V}_{0\mu} + g_0 v_{0\mu}.
\label{eq:A9}
\end{equation}

Expanding the ${\cal O}(p^2)$ Lagrangian Eq.(\ref{eq:22p2}) in terms
of these quantum fields $u$ and $v_\mu$, we obtain
\begin{equation}
  {\cal L}_2 = \left. {\cal L}_2 \right|_0 
   + \left. {\cal L}_2 \right|_u + \left. {\cal L}_2 \right|_v 
   + \left. {\cal L}_2 \right|_{uu} + \left. {\cal L}_2 \right|_{vv} 
   +\left. {\cal L}_2 \right|_{uv} + \cdots,
\end{equation}
with
\begin{eqnarray}
  \left.  {\cal L}_2 \right|_0
  &=& \dfrac{f^2}{4} \tr\left[
    (\bar{V}_{0\mu} -\bar{V}_{1\mu} )
    (\bar{V}_{0}^{\mu} -\bar{V}_{1}^{\mu} )
    \right]
   -\dfrac{1}{2 g_0^2} 
    \tr\left[\bar{V}_{0\mu\nu}\bar{V}_0^{\mu\nu} \right] 
   -\dfrac{1}{2 g_1^2} 
    \tr\left[\bar{V}_{1\mu\nu}\bar{V}_1^{\mu\nu} \right],
  \\
  \left.  {\cal L}_2 \right|_u
  &=& \dfrac{f}{2} \partial_\mu u^a 
      (\bar{V}_0^{a\mu} - \bar{V}_1^{a\mu}) 
     -\dfrac{f}{2} \epsilon^{abc} 
      \bar{V}_{0\mu}^{a} \bar{V}_1^{b\mu} u^c,
\label{eq:lag1_u}
  \\
  \left. {\cal L}_2 \right|_v
  &=& \dfrac{f^2}{4} (\bar{V}_{0\mu}^{a}-\bar{V}_{1\mu}^{a})
      (g_0 v_0^\mu - g_1 v_1^\mu)
     -\frac{1}{g_0} \bar{V}_{0\mu\nu}^a (D^\mu v_0^\nu)^a
     -\frac{1}{g_1} \bar{V}_{1\mu\nu}^a (D^\mu v_1^\nu)^a,
\label{eq:lag1_v}
  \\
  \left. {\cal L}_2 \right|_{uu}
  &=& \frac{1}{2} (D_\mu u)^a (D^\mu u)^a
     -\frac{1}{8} \epsilon^{abc} \epsilon^{ade}
      (\bar{V}_{0\mu}^b - \bar{V}_{1\mu}^b) u^c
      (\bar{V}_{0}^{d\mu} - \bar{V}_{1}^{d\mu}) u^e,
  \\
  \left. {\cal L}_2 \right|_{vv}
  &=& \dfrac{f^2}{8} (g_0 v_{0\mu}^a - g_1 v_{1\mu}^a )
                     (g_0 v_{0}^{a\mu} - g_1 v_{1}^{a\mu} )
  \nonumber\\
  & &
     -\frac{1}{2}(D_\mu v_{0\nu})^a (D^\mu v_{0}^{\nu})^a 
     +\frac{1}{2}(D_\mu v_{0\nu})^a (D^\nu v_{0}^{\mu})^a 
     +\frac{1}{2}\epsilon^{abc} 
      \bar{V}_{0\mu\nu}^a v_0^{b\mu}v_0^{c\nu}
  \nonumber\\
  & &
     -\frac{1}{2}(D_\mu v_{1\nu})^a (D^\mu v_{1}^{\nu})^a 
     +\frac{1}{2}(D_\mu v_{1\nu})^a (D^\nu v_{1}^{\mu})^a 
     +\frac{1}{2}\epsilon^{abc} 
      \bar{V}_{1\mu\nu}^a v_1^{b\mu}v_1^{c\nu},
  \\
  \left. {\cal L}_2 \right|_{uv}
  &=& \frac{f}{2} (\partial_\mu u^a) (g_0 v_0^{a\mu} - g_1 v_1^{a\mu})
     -\frac{f}{2} g_0 \epsilon^{abc} u^a v_{0}^{b\mu} \bar{V}_{1\mu}^c
     +\frac{f}{2} g_1 \epsilon^{abc} u^a v_{1}^{b\mu} \bar{V}_{0\mu}^c.
\end{eqnarray}
Here $\bar{V}_{0\mu\nu}$ and $\bar{V}_{1\mu\nu}$ are defined as
\begin{equation}
  \bar{V}_{0\mu\nu} \equiv \partial_\mu \bar{V}_{0\nu} 
                         - \partial_\nu \bar{V}_{0\mu} 
                    +i[ \bar{V}_{0\mu}, \bar{V}_{0\nu}], \qquad
  \bar{V}_{1\mu\nu} \equiv \partial_\mu \bar{V}_{1\nu} 
                         - \partial_\nu \bar{V}_{1\mu} 
                    +i[ \bar{V}_{1\mu}, \bar{V}_{1\nu}].
\end{equation}
We also define the  covariant derivatives $D_\mu u$, $D_\mu v_{0\nu}$,
and  $D_\mu v_{1\nu}$, by
\begin{eqnarray}
  (D_\mu u)^a 
  &\equiv& \partial_\mu u^a - \frac{1}{2} \epsilon^{abc} 
           (\bar{V}_{0\mu}^b + \bar{V}_{1\mu}^b ) u^c ,  
\label{eq:covariant_derivative}
  \\
  (D_\mu v_{0\nu})^a
  &\equiv& \partial_\mu v_{0\nu}^a 
         - \epsilon^{abc} \bar{V}_{0\mu}^b v_{0\nu}^c, 
  \\
  (D_\mu v_{1\nu})^a
  &\equiv& \partial_\mu v_{1\nu}^a 
         - \epsilon^{abc} \bar{V}_{1\mu}^b v_{1\nu}^c . 
\end{eqnarray}
Note that, from Eqn.~(\ref{eq:covariant_derivative}), the fluctuating
pion fields transform as adjoints of the unbroken diagonal subgroup. However,
with respect to either gauged $SU(2)$, the coupling of the pions to $\bar{V}^b_{0\mu}$
or $\bar{V}^b_{1\mu}$ is precisely {\it one-half} of the value for an adjoint scalar field.

In order to compute radiative corrections, we
introduce the background field gauge fixing Lagrangian,
\begin{equation}
  {\cal L}_{\rm GF} =
    -\frac{1}{2\xi} 
     \left((D_\mu v_0^\mu)^a - \xi \dfrac{g_0 f}{2} u^a\right)^2
    -\frac{1}{2\xi} 
     \left((D_\mu v_1^\mu)^a + \xi \dfrac{g_1 f}{2} u^a\right)^2~,
\label{eq:GF1}
\end{equation}
for the quantum fields.
We then obtain
\begin{eqnarray}
  \left. {\cal L}_2 \right|_{uu} + \left. {\cal L}_{\rm GF} \right|_{uu}
  &=& \frac{1}{2} (D_\mu u)^a (D^\mu u)^a
     -\frac{1}{8} \epsilon^{abc} \epsilon^{ade}
      (\bar{V}_{0\mu}^b - \bar{V}_{1\mu}^b) u^c
      (\bar{V}_{0}^{d\mu} - \bar{V}_{1}^{d\mu}) u^e
  \nonumber\\
  & & -\xi \dfrac{(g_0^2+g_1^2)f^2}{8} u^a u^a~,
\label{eq:lag1_uu}
  \\
  \left. {\cal L}_2 \right|_{vv} + \left. {\cal L}_{\rm GF}
  \right|_{vv}
  &=& \dfrac{f^2}{8} (g_0 v_{0\mu}^a - g_1 v_{1\mu}^a )
                     (g_0 v_{0}^{a\mu} - g_1 v_{1}^{a\mu} )
  \nonumber\\
  & &-\frac{1}{2}(D_\mu v_{0\nu})^a (D^\mu v_{0}^{\nu})^a 
     +\frac{1}{2}\left(1-\frac{1}{\xi}\right)
      (D_\mu v_{0}^{\mu})^a (D_\nu v_{0}^{\nu})^a 
  \nonumber\\
  & &
     +\epsilon^{abc} \bar{V}_{0\mu\nu}^a v_0^{b\mu}v_0^{c\nu}
  \nonumber\\
  & &-\frac{1}{2}(D_\mu v_{1\nu})^a (D^\mu v_{1}^{\nu})^a 
     +\frac{1}{2}\left(1-\frac{1}{\xi}\right)
      (D_\mu v_{1}^{\mu})^a (D_\nu v_{1}^{\nu})^a 
  \nonumber\\
  & &
     +\epsilon^{abc} \bar{V}_{1\mu\nu}^a v_1^{b\mu}v_1^{c\nu},
\label{eq:lag1_vv}
  \\
  \left. {\cal L}_2 \right|_{uv} + \left. {\cal L}_{\rm GF} \right|_{uv}
  &=& -\frac{f}{2} \epsilon^{abc} u^a 
       \left(\bar{V}_{0\mu}^b - \bar{V}_{1\mu}^b  \right)
       \left(g_0 v_0^{c\mu} + g_1 v_1^{c\mu}\right),
\label{eq:lag1_uv}
\end{eqnarray}
up to terms proportional to total derivatives.

We also need to introduce the Faddeev-Popov Lagrangian associated with 
the gauge fixing term Eq.(\ref{eq:GF1}),
\begin{eqnarray}
  {\cal L}_{\rm FP} &=& 
   (D_\mu \bar{c}_0)^a 
   ((D^\mu c_0)^a - \epsilon^{abc} v_{0}^{b\mu} c_0^c)
  +(D_\mu \bar{c}_1)^a 
   ((D^\mu c_1)^a - \epsilon^{abc} v_{1}^{b\mu} c_1^c)
  \nonumber\\
  & & -\xi\dfrac{f^2}{4} (g_0 \bar{c}_0^a - g_1 \bar{c}_1^a)
       (g_0 c_0^a - g_1 c_1^a) + \cdots,
\label{eq:lag1_FP}
\end{eqnarray}
with $c_0$, $c_1$ ($\bar{c}_0$, $\bar{c}_1$) being the Faddeev-Popov
ghost (anti-ghost) for the site-0,1 gauge fields.   Here $D_\mu c$ is
defined as
\begin{equation}
  (D_\mu \bar{c}_0)^a 
  = \partial_\mu \bar{c}_0^a - \epsilon^{abc} \bar{V}_{0\mu}^b \bar{c}_0^c,
  \qquad
  (D_\mu {c}_0)^a 
  = \partial_\mu {c}_0^a - \epsilon^{abc} \bar{V}_{0\mu}^b {c}_0^c,
\end{equation}
\begin{equation}
  (D_\mu \bar{c}_1)^a 
  = \partial_\mu \bar{c}_1^a - \epsilon^{abc} \bar{V}_{1\mu}^b \bar{c}_1^c,
  \qquad
  (D_\mu {c}_1)^a 
  = \partial_\mu {c}_1^a - \epsilon^{abc} \bar{V}_{1\mu}^b {c}_1^c.
\end{equation}
In Eq.(\ref{eq:lag1_FP}) we 
do not specify the  terms of higher order in the fluctuating pion fields
arising from the {\it nonlinear} gauge transformation properties of
these fields, terms 
of the form $u^n\bar{c}c$ terms ($n \ge 1$).
For the one-loop analysis it is enough to consider just the $\bar{c}c$ terms
in this  Faddeev-Popov Lagrangian.

The Lagrangians Eq.(\ref{eq:lag1_u}) and Eq.(\ref{eq:lag1_v}) lead to
the following equations of motion for the background fields,
\begin{equation}
  0 = \partial_\mu (\bar{V}_0^{a\mu} - \bar{V}_1^{a\mu})
     +\epsilon^{abc} \bar{V}_{0\mu}^{b} \bar{V}_1^{c\mu}
\end{equation}
and
\begin{eqnarray}
  & & 0 = \partial_\mu \bar{V}_0^{a\mu\nu}
         -\epsilon^{abc} \bar{V}_{0\mu}^b \bar{V}_0^{c\mu\nu}
         +\dfrac{g_0^2 f^2}{4} 
          (\bar{V}_0^{a\nu} - \bar{V}_1^{a\nu}), 
  \\
  & & 0 = \partial_\mu \bar{V}_1^{a\mu\nu}
         -\epsilon^{abc} \bar{V}_{1\mu}^b \bar{V}_1^{c\mu\nu}
         -\dfrac{g_1^2 f^2}{4} 
          (\bar{V}_0^{a\nu} - \bar{V}_1^{a\nu}). 
\end{eqnarray}
As is usual for the background field method, we assume that the background
fields satisfy these equations of motion, so that quantum corrections
arise only from $uu$, $uv$, $vv$, and $cc$ loops.

\DOUBLEFIGURE[tb]{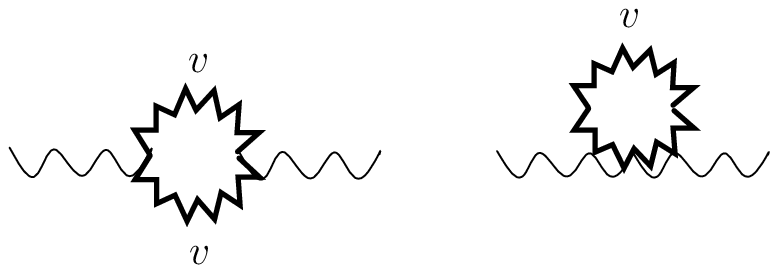,width=7cm}{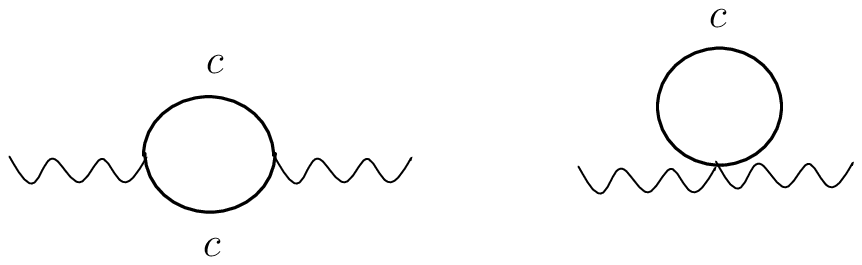,width=7cm}{Gauge boson loop diagrams.\label{fig:vvloops}}{Ghost loop diagrams.\label{fig:ccloops}}

\subsubsection{Gauge boson loop ($vv$)}

Now we are ready to evaluate the one-loop divergences of the
$SU(2)\times SU(2)$ gauged nonlinear sigma model.
We first consider the gauge boson loop diagrams ($vv$ loop diagrams)
arising from Eq.~(\ref{eq:lag1_vv}), as illustrated in Fig. \ref{fig:vvloops}.
In Landau gauge\footnote{The behavior proportional to $1/\xi$ in 
Eq.~(\protect\ref{eq:lag1_vv}) is cancelled by the $\xi q^\mu q^\nu$ term in the
gauge-boson propagator, leading to a smooth $\xi \to 0$ limit.}  ($\xi=0$), 
we find the effective Lagrangian
generated from $vv$ diagrams,
\begin{equation}
   \frac{20}{3}
   \dfrac{1}{(4\pi)^2\bar{\epsilon}}
   \tr\left[\bar{V}_{0\mu\nu}\bar{V}_0^{\mu\nu} \right] 
 +\frac{20}{3}
   \dfrac{1}{(4\pi)^2\bar{\epsilon}}
   \tr\left[\bar{V}_{1\mu\nu}\bar{V}_1^{\mu\nu} \right].
\label{eq:efflag1_vv}
\end{equation}
Here $\bar{\epsilon}$ is defined as
\begin{equation}
  \dfrac{1}{\bar{\epsilon}} \equiv 
  \dfrac{\Gamma(2-d/2)}{2(4\pi)^{d/2-2}},
\end{equation}
with $d$ being the dimensionality of space-time.

\subsubsection{Ghost loop ($cc$)}

We next consider effects of ghost loop ($cc$ diagrams).
In Landau gauge ($\xi=0$) the ghosts remain massless.
From the Lagrangian Eq.(\ref{eq:lag1_FP}), we find
the  effective Lagrangian arising from the $cc$ diagrams illustrated
in Fig. \ref{fig:ccloops},
\begin{equation}
   \frac{2}{3}
   \dfrac{1}{(4\pi)^2\bar{\epsilon}}
   \tr\left[\bar{V}_{0\mu\nu}\bar{V}_0^{\mu\nu} \right] 
  +\frac{2}{3}
   \dfrac{1}{(4\pi)^2\bar{\epsilon}}
   \tr\left[\bar{V}_{1\mu\nu}\bar{V}_1^{\mu\nu} \right].
\label{eq:efflag1_cc}
\end{equation}

\DOUBLEFIGURE[tb]{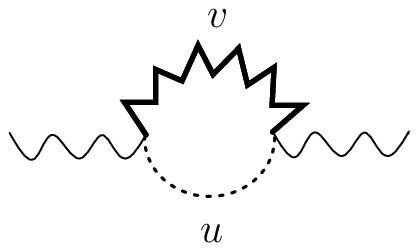,width=3cm}{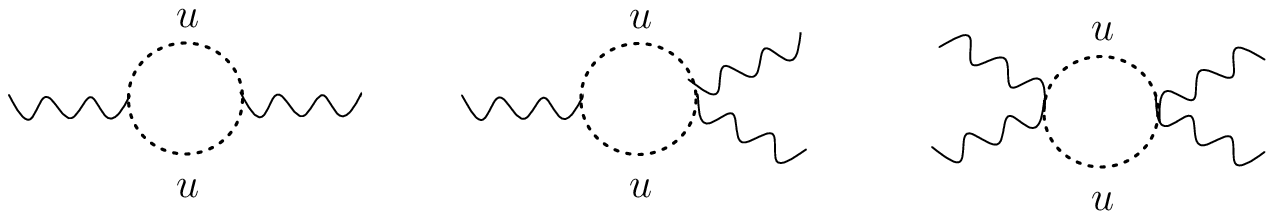,width=10cm}{Gauge-NGB loop diagram.\label{fig:uvloop}}{NGB loop diagrams.\label{fig:uuloops}}

\subsubsection{Gauge-NGB mixed loop ($uv$)}

We next evaluate the gauge-boson and Nambu-Goldstone-boson mixed loop
diagrams ($uv$ diagrams) arising from Eq.(\ref{eq:lag1_uv}), illustrated
in Fig. \ref{fig:uvloop}.  
In Landau gauge we find the 
$uv$-generated one-loop effective Lagrangian,
\begin{equation}
  \dfrac{-3}{(4\pi)^2\bar{\epsilon}}(g_0^2 + g_1^2) \dfrac{f^2}{4} 
  \tr\left[
    (\bar{V}_{0\mu} -\bar{V}_{1\mu} )
    (\bar{V}_{0}^{\mu} -\bar{V}_{1}^{\mu} )
    \right].
\label{eq:efflag1_uv}
\end{equation}

\subsubsection{NGB loop ($uu$)}

Finally, we turn to the Nambu-Goldstone-boson loop effects ($uu$ loop)
generated from the Lagrangian Eq.(\ref{eq:lag1_uu}), illustrated
in Fig. \ref{fig:uuloops}.
In the Landau gauge ($\xi=0$), the
Nambu-Goldstone bosons remain massless, and 
we are thus able to use the result of conventional chiral perturbation
theory.  See also Appendix  \ref{sec:divergences} of this note.
We find that the $uu$ diagrams lead to the one-loop effective action,
\begin{eqnarray}
  & &
  -\frac{1}{12} \dfrac{1}{(4\pi)^2 \bar{\epsilon}} \tr\left[
    \bar{V}_{0\mu\nu} \bar{V}_{0}^{\mu\nu} 
   \right]
  -\frac{1}{12} \dfrac{1}{(4\pi)^2 \bar{\epsilon}} \tr\left[
    \bar{V}_{1\mu\nu} \bar{V}_{1}^{\mu\nu} 
   \right]
  -\frac{1}{6} \dfrac{1}{(4\pi)^2 \bar{\epsilon}} \tr\left[
    \bar{V}_{0\mu\nu} \bar{V}_{1}^{\mu\nu} 
   \right]
  \nonumber\\
  & &
  +\frac{1}{6} \dfrac{1}{(4\pi)^2 \bar{\epsilon}} i \tr\left[
    (\bar{V}_{0\mu}-\bar{V}_{1\mu}) 
    (\bar{V}_{0\nu}-\bar{V}_{1\nu}) 
    \bar{V}_{0}^{\mu\nu} 
   \right]
  +\frac{1}{6} \dfrac{1}{(4\pi)^2 \bar{\epsilon}} i \tr\left[
    (\bar{V}_{0\mu}-\bar{V}_{1\mu}) 
    (\bar{V}_{0\nu}-\bar{V}_{1\nu}) 
    \bar{V}_{1}^{\mu\nu} 
   \right]
  \nonumber\\
  & &
  +\frac{1}{6}\dfrac{1}{(4\pi)^2 \bar{\epsilon}} \tr\left[
    (\bar{V}_{0\mu}-\bar{V}_{1\mu}) 
    (\bar{V}_{0\nu}-\bar{V}_{1\nu}) 
  \right] \tr\left[
    (\bar{V}_{0}^{\mu}-\bar{V}_{1}^{\mu}) 
    (\bar{V}_{0}^{\nu}-\bar{V}_{1}^{\nu}) 
  \right]
  \nonumber\\
  & &
  +\frac{1}{12}\dfrac{1}{(4\pi)^2 \bar{\epsilon}} \tr\left[
    (\bar{V}_{0\mu}-\bar{V}_{1\mu}) 
    (\bar{V}_{0}^{\mu}-\bar{V}_{1}^{\mu}) 
  \right] \tr\left[
    (\bar{V}_{0\nu}-\bar{V}_{1\nu}) 
    (\bar{V}_{0}^{\nu}-\bar{V}_{1}^{\nu}) 
  \right].
\label{eq:efflag1_uu}
\end{eqnarray}
In order to renormalize the  one-loop divergences in Eq.(\ref{eq:efflag1_uu}),
we need to introduce counter terms at ${\cal O}(p^4)$,
\begin{eqnarray}
  {\cal L}_{4}
  &=& \alpha_{1} \tr\left[ 
        V_{0\mu\nu} U V_{1}^{\mu\nu} U^\dagger
      \right]
  \nonumber\\
  & & -2i\alpha_{2} \tr\left[
        (D_\mu U)^\dagger (D_\nu U) V_{1\mu\nu}
      \right]
  \nonumber\\
  & & -2i\alpha_{3} \tr\left[
        V_{0}^{\mu\nu} (D_\mu U) (D_\nu U)^\dagger
      \right] 
  \nonumber\\
  & & +\alpha_{4} 
       \tr\left[(D_\mu U) (D_\nu U)^\dagger \right] 
       \tr\left[(D^\mu U) (D^\nu U)^\dagger \right] 
  \nonumber\\
  & & +\alpha_{5} 
       \tr\left[(D_\mu U) (D^\mu U)^\dagger \right] 
       \tr\left[(D_\nu U) (D^\nu U)^\dagger \right],
\end{eqnarray}
which, using Eqs.~(\ref{eq:A8}) and (\ref{eq:A9}), are equivalent to the forms of the interactions
in Eq.~(\ref{eq:efflag1_uu}).

\subsubsection{$SU(2) \times SU(2)$ Renormalization group equations}

We next define the $\overline{\rm MS}$ renormalized parameters.
From Eq.(\ref{eq:efflag1_vv}), Eq.(\ref{eq:efflag1_cc})
and  Eq.(\ref{eq:efflag1_uu}), we define ``renormalized'' gauge
coupling strengths $g_{0r}$ and $g_{1r}$ by,
\begin{eqnarray}
  \dfrac{1}{g_{0r}^2} 
  &=& \dfrac{1}{g_0^2} -\dfrac{1}{(4\pi)^2 \bar{\epsilon}} \left(
    \dfrac{40}{3} + \dfrac{4}{3} - \dfrac{1}{6} \right),
  \\
  \dfrac{1}{g_{1r}^2} 
  &=& \dfrac{1}{g_1^2} -\dfrac{1}{(4\pi)^2 \bar{\epsilon}} \left(
    \dfrac{40}{3} + \dfrac{4}{3} - \dfrac{1}{6} \right).
\end{eqnarray}
Renormalization of the decay constant $f$ is given by
\begin{equation}
  f_r^2 
  = f^2 -3(g_0^2+g_1^2) f^2 \dfrac{1}{(4\pi)^2 \bar{\epsilon}}, 
\end{equation}
where we used the result of $uv$-loop diagrams
Eq.(\ref{eq:efflag1_uv}).
The renormalization of the ${\cal O}(p^4)$ Lagrangian can be determined
from Eq.(\ref{eq:efflag1_uu}).  
We find
\begin{eqnarray}
  \alpha_1^r &=& 
  \alpha_1 - \dfrac{1}{6} \dfrac{1}{(4\pi)^2 \bar{\epsilon}}, \\
  \alpha_2^r &=& 
  \alpha_2 - \dfrac{1}{12} \dfrac{1}{(4\pi)^2 \bar{\epsilon}}, \\
  \alpha_3^r &=& 
  \alpha_3 - \dfrac{1}{12} \dfrac{1}{(4\pi)^2 \bar{\epsilon}}, \\
  \alpha_4^r &=& 
  \alpha_4 + \dfrac{1}{6} \dfrac{1}{(4\pi)^2 \bar{\epsilon}}, \\
  \alpha_5^r &=& 
  \alpha_5 + \dfrac{1}{12} \dfrac{1}{(4\pi)^2 \bar{\epsilon}}.
\end{eqnarray}

It is  now straightforward to obtain the $\overline{\rm MS}$
renormalization group equations of these parameters.
We find
\begin{eqnarray}
  \mu \dfrac{d}{d\mu} \left(\dfrac{1}{g_{0r}^2}\right)
  &=& \dfrac{1}{(4\pi)^2} 
      \left[ \dfrac{44}{3} - \dfrac{1}{6} \right],
\label{eq:rge1_g0r}
  \\
  \mu \dfrac{d}{d\mu} \left(\dfrac{1}{g_{1r}^2}\right)
  &=& \dfrac{1}{(4\pi)^2} 
      \left[ \dfrac{44}{3} - \dfrac{1}{6} \right],
\label{eq:rge1_g1r}
\end{eqnarray}
for the renormalized gauge coupling strengths.
Here $\mu$ stands for the $\overline{\rm MS}$ renormalization scale.
Note that the factor $-1/6$ in Eqs.(\ref{eq:rge1_g0r}) and
(\ref{eq:rge1_g1r}) comes from the Nambu-Goldstone boson loop
($uu$-diagrams).
This factor is a quarter the size of the usual $SU(2)$ adjoint scalar-loop
effect in the gauge coupling renormalization group equations ($-2/3$),
with the 
difference arising from the definition of ``covariant'' derivative
Eq.(\ref{eq:covariant_derivative}) for the $u$-field.
The renormalization group running of the $f$-constant comes from the
$uv$-loop diagram in Landau gauge, and we find
\begin{eqnarray}
  \mu \dfrac{d}{d\mu} f_r^2 
    &=& \dfrac{3}{(4\pi)^2} (g_{0r}^2 + g_{1r}^2) f_r^2. 
\end{eqnarray}

Finally, we compute the renormalization group equations for
the ${\cal O}(p^4)$ interactions
\begin{eqnarray}
  \mu \dfrac{d}{d\mu} \alpha_{1}^r
  &=& \dfrac{1}{6(4\pi)^2}, 
  \\
  \mu \dfrac{d}{d\mu} \alpha_{2}^r 
  &=& \dfrac{1}{12(4\pi)^2}, 
  \\
  \mu \dfrac{d}{d\mu} \alpha_{3}^r 
  &=& \dfrac{1}{12(4\pi)^2}, 
  \\
  \mu \dfrac{d}{d\mu} \alpha_{4}^r 
  &=& -\dfrac{1}{6(4\pi)^2}, 
  \\
  \mu \dfrac{d}{d\mu} \alpha_{5}^r 
  &=& -\dfrac{1}{12(4\pi)^2}. 
\end{eqnarray}
It should also be emphasized that the RGE equations for the
${\cal O}(p^4)$ operators in the gauged non-linear sigma model are,
at this order, 
{\it identical} with those in the usual global (non-gauged) sigma model.

\subsubsection{Delocalization operator}

\EPSFIGURE{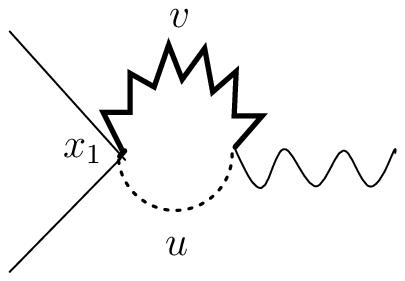,width=3cm}{Gauge-NGB loop diagram responsible
for the renormalization of $x_1$.\label{fig:x1graph}}

We next consider the renormalization of the
delocalization operator,
\begin{equation}
  x_1 \tr\left[ J^\mu D_\mu U U^\dagger \right].
\label{eq:x1_op}
\end{equation}
Expanding Eq.(\ref{eq:x1_op}) in terms of $u$ and $v_\mu$, we find
\begin{equation}
  \left. x_1  \tr\left[ J^\mu D_\mu U U^\dagger \right]\right|_0
  = x_1 i \tr\left[ 
        J^\mu \bar{U} (\bar{V}_{0\mu} - \bar{V}_{1\mu}) \bar{U}^\dagger 
      \right], 
\label{eq:x1_0}
\end{equation}
\begin{equation}
  \left. x_1 \tr\left[ J^\mu D_\mu U U^\dagger \right]\right|_{uv}
  = -2 x_1 \dfrac{g_1}{f} i \epsilon^{abc} v_{1\mu}^b u^c 
     \tr\left[ J^\mu \bar{U} T^a \bar{U}^\dagger\right].
\label{eq:x1_uv}
\end{equation}
The $uv$ loop diagram arising from Eq.(\ref{eq:lag1_uv}) and
Eq.(\ref{eq:x1_uv}) then leads (see Fig. \ref{fig:x1graph}) 
to the one-loop effective operator,
\begin{equation}
  -\dfrac{3g_1^2}{(4\pi)^2 \bar{\epsilon}} x_1 i \tr\left[
    J^\mu \bar{U} (\bar{V}_{0\mu} - \bar{V}_{1\mu}) \bar{U}^\dagger 
   \right], 
\label{eq:eff_x1}
\end{equation}
in Landau gauge.
Comparing Eq.(\ref{eq:x1_0}) and Eq.(\ref{eq:eff_x1}), we find the 
renormalization of the delocalization parameter $x_1$,
\begin{equation}
  x_1^r = x_1  - \dfrac{3g_1^2}{(4\pi)^2 \bar{\epsilon}} x_1,
\end{equation}
and therefore the renormalization group equation for  $x_1$
\begin{equation}
  \mu \dfrac{d}{d\mu} x_1^r = \dfrac{3g_{1r}^2}{(4\pi)^2} x_1^r.
\end{equation}

\subsection{$SU(2)\times U(1)$ gauged nonlinear sigma model}
\label{sec:21model}

\EPSFIGURE[tb]{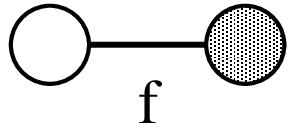,width=4cm}
 {Moose diagram \protect\cite{Georgi:1985hf}  for the two site electroweak chiral Lagrangian, an
  $SU(2)\times U(1)$ gauged nonlinear sigma model. The $SU(2)_1$ gauge group is shown
  as an open circle; the $U(1)_2$ gauge group as a shaded circle. The link represents the nonlinear
  sigma model field $U$, with $f$-constant $f$.
 \label{fig:twosite2}}
 
We next discuss an $SU(2)\times U(1)$ gauged nonlinear sigma model
of the sort shown in Figure~\ref{fig:twosite2}.
The lowest order (${\cal O}(p^2)$) Lagrangian of this model is given
by
\begin{eqnarray}
  {\cal L}_2 
  &=& 
    \dfrac{f^2}{4} 
    \tr\left[(D_\mu U)^\dagger (D^\mu U) \right]
  +\beta\dfrac{f^2}{4} 
      \tr\left[  U^\dagger (D_\mu U) \tau_3 \right]
      \tr\left[  U^\dagger (D^\mu U) \tau_3 \right]
  \nonumber\\
  & &
   -\dfrac{1}{2g_1^2}
    \tr\left[V_{1\mu\nu}V_1^{\mu\nu} \right]
   -\dfrac{1}{2g_2^2}
    \tr\left[V_{2\mu\nu}V_2^{\mu\nu} \right],
\label{eq:21p2}
\end{eqnarray}
with $U$ being a chiral field,
\begin{equation}
  U \equiv \exp \left[ \dfrac{2i\pi^a T^a}{f} \right], \qquad
  T^a \equiv \dfrac{\tau^a}{2}. 
\end{equation}
The $SU(2)_1$ and $U(1)_2$ gauge fields (at sites 1 and 2 in the
moose diagram Figure~\ref{fig:twosite2}) are 
\begin{equation}
  V_{1\mu} \equiv V^a_{1\mu} T^a, \qquad
  V_{2\mu} \equiv V^3_{2\mu} T^3.
\end{equation}
The covariant derivative $D_\mu U$ is given by
\begin{equation}
  D_\mu U \equiv \partial_\mu U + i V_{1\mu} U - i U V_{2\mu}.
\end{equation}
Note that the $U(1)_2$ gauge group is embedded as the $T_3$ generator
of $SU(2)_2$ in this Lagrangian.
The term proportional to $\beta$ in Eq.(\ref{eq:21p2}) keeps the
$U(1)_2$ invariance, but it violates $SU(2)_2$, and hence ``custodial'' symmetry
as well.  
The gauge field strengths $V_{1\mu\nu}$ and $V_{2\mu\nu}$ are defined
as
\begin{equation}
  V_{1\mu\nu} \equiv \partial_\mu V_{1\nu} - \partial_\nu V_{1\mu} 
                    +i[ V_{1\mu}, V_{1\nu}], \qquad
  V_{2\mu\nu} \equiv \partial_\mu V_{2\nu} - \partial_\nu V_{2\mu}. 
\end{equation}

We calculate one-loop divergences arising from Eq.(\ref{eq:21p2})
using methods similar to \S\ref{sec:22model}.
We decompose the chiral field $U$ into a background field $\bar{U}$
and fluctuating quantum fields $u^1, u^2$ and $u_z$,
\begin{equation}
  U = \bar{U} \exp\left[\dfrac{2i(u^1 T^1 + u^2 T^2)}{f} \right]
              \exp\left[\dfrac{2i u_z T^3}{f_z}\right].
\end{equation}
Here we allow for  the neutral current $f$-constant, $f_z$,
to differ from $f$
\begin{equation}
  \dfrac{f_z^2}{f^2} \equiv 1 - 2\beta.
\label{eq:f_z}
\end{equation}
The site-2 gauge field $V_2^\mu$ is decomposed into background and fluctuating
fields,
\begin{equation}
  V_{2\mu}^{3} = \bar{V}_{2\mu}^{3} + g_2 v_{2\mu}^{3}~,
\end{equation}
as is the site-1 gauge field $V_1^\mu$
\begin{equation}
  V_{1\mu}' = \bar{U}^\dagger V_{1\mu} \bar{U} 
                - i \bar{U}^\dagger \partial_\mu \bar{U}
             = \bar{V}_{1\mu} + g_1 v_{1\mu}.
\end{equation}
We introduce the gauge fixing Lagrangian,
\begin{eqnarray}
  {\cal L}_{\rm GF} &=& 
  -\frac{1}{2\xi}\left[ 
    (D_\mu v_1^\mu)^1 - \xi g_1 \frac{f}{2} u^1
  \right]^2
  - 
  \frac{1}{2\xi}\left[ 
    (D_\mu v_1^\mu)^2 - \xi g_1 \frac{f}{2} u^2
  \right]^2 \nonumber\\
  & &
  -
  \frac{1}{2\xi}\left[ 
    (D_\mu v_1^\mu)^3 - \xi g_1 \frac{f_z}{2} u_z
  \right]^2 
  -
  \frac{1}{2\xi}\left[ 
    \partial_\mu v_2^{3\mu} + \xi g_2 \frac{f_z}{2} u_z
  \right]^2, 
\end{eqnarray}
where
\begin{equation}
  D_\mu v_1^\mu 
  = \left(\begin{array}{l}
      \partial_\mu v_1^{1\mu} - \bar{V}_{1\mu}^2 v_1^{3\mu} 
                              + \bar{V}_{1\mu}^3 v_1^{2\mu} \\
      \partial_\mu v_1^{2\mu} - \bar{V}_{1\mu}^3 v_1^{1\mu} 
                              + \bar{V}_{1\mu}^1 v_1^{3\mu} \\
      \partial_\mu v_1^{3\mu} - \bar{V}_{1\mu}^1 v_1^{2\mu} 
                              + \bar{V}_{1\mu}^2 v_1^{1\mu}
    \end{array}\right).
\label{eq:GF2}
\end{equation}

Expanding these Lagrangians Eq.(\ref{eq:21p2}) and Eq.(\ref{eq:GF2})
in terms of the fluctuating quantum fields $u$ and $v^\mu$, we find
\begin{equation}
  \left. {\cal L}_2 \right|_{uu}
 +\left. {\cal L}_{\rm GF} \right|_{uu}
  = \frac{1}{2} (D_\mu u)^T (D^\mu u) - \frac{1}{2} u^T \sigma u,
\label{eq:lag2_uu}
\end{equation}
with
\begin{eqnarray}
  u \equiv \left(\begin{array}{c} u^1 \\ u^2 \\ u_z
        \end{array}\right)~,  \qquad \qquad
  D_\mu u \equiv \partial_\mu u + \Gamma_\mu u ~,
\end{eqnarray}
\begin{equation}
 \Gamma_\mu = \left(\begin{array}{ccc}
       0  
     & \dfrac{1}{2} (2-\dfrac{f_z^2}{f^2}) \bar{V}_{1\mu}^3
       +\dfrac{1}{2} \dfrac{f_z^2}{f^2} \bar{V}_{2\mu}^3 
     & -\dfrac{f_z}{2f} \bar{V}_{1\mu}^2
     \\
       -\dfrac{1}{2} (2-\dfrac{f_z^2}{f^2}) \bar{V}_{1\mu}^3
       -\dfrac{1}{2} \dfrac{f_z^2}{f^2} \bar{V}_{2\mu}^3 
     & 0  
     & \dfrac{f_z}{2f}  \bar{V}_{1\mu}^1
     \\
       \dfrac{f_z}{2f}  \bar{V}_{1\mu}^2
     & -\dfrac{f_z}{2f} \bar{V}_{1\mu}^1
     & 0
                       \end{array}\right),
\end{equation}
\begin{eqnarray}
  \sigma_{11} &\equiv&
     \dfrac{1}{4}(4-3\frac{f_z^2}{f^2}) 
     \bar{V}_{1\mu}^2 \bar{V}_{1}^{2\mu}
    +\dfrac{1}{4} \dfrac{f_z^4}{f^4} 
     (\bar{V}_{1\mu}^3 - \bar{V}^3_{2\mu})
     (\bar{V}_{1}^{3\mu} - \bar{V}_{2}^{3\mu})
    +\xi\dfrac{g_1^2 f^2}{4},
  \\
  \sigma_{12} &\equiv&
    -\frac{1}{4}(4-3\frac{f_z^2}{f^2}) \bar{V}_{1\mu}^1
                                       \bar{V}_{1}^{2\mu},
  \\
  \sigma_{1z} &\equiv&
    -\dfrac{1}{4} \dfrac{f_z^3}{f^3} 
     \bar{V}_{1\mu}^1 (\bar{V}_{1}^{3\mu} - \bar{V}_{2}^{3\mu}), 
  \\
  \sigma_{22} &\equiv&
     \dfrac{1}{4}(4-3\dfrac{f_z^2}{f^2}) \bar{V}_{1\mu}^1 \bar{V}_{1}^{1\mu}
    +\dfrac{1}{4} \dfrac{f_z^4}{f^4} 
     (\bar{V}_{1\mu}^3 -\bar{V}_{2\mu}^3)
     (\bar{V}_{1}^{3\mu} -\bar{V}_{2}^{3\mu})
    +\xi\dfrac{g_1^2 f^2}{4},
  \\
  \sigma_{2z} &\equiv&
    -\frac{1}{4} \dfrac{f_z^3}{f^3} 
     \bar{V}_{1\mu}^2 (\bar{V}_1^{3\mu} - \bar{V}_2^{3\mu})~,
  \\
  \sigma_{zz} &\equiv&
    \dfrac{f_z^2}{4f^2} (\bar{V}_{1\mu}^1 \bar{V}_1^{1\mu}
                       +\bar{V}_{1\mu}^2 \bar{V}_1^{2\mu})
   +\xi \dfrac{(g_1^2 + g_2^2) f_z^2}{4} ,
\end{eqnarray}
where we have simplified these expressions by using the equations of motion of the background field. 
We also find
\begin{eqnarray}
  \left. {\cal L}_2 \right|_{uv}
 +\left. {\cal L}_{\rm GF} \right|_{uv}
  &=& - g_1 \dfrac{f}{2} (2-\dfrac{f_z^2}{f^2}) 
          \left(
             \bar{V}_{1\mu}^2 v_1^{3\mu} u^1 
           - \bar{V}_{1\mu}^1 v_1^{3\mu} u^2
          \right)
  \nonumber\\
  & &  -g_2 \dfrac{f_z^2}{2f} 
          \left(
             \bar{V}_{1\mu}^2 v_2^{3\mu} u^1 
           - \bar{V}_{1\mu}^1 v_2^{3\mu} u^2
          \right)
  \nonumber\\
  & &
      - g_1 \dfrac{f_z^2}{2f} \left(
          (\bar{V}_{1\mu}^3-\bar{V}_{2\mu}^3) v_1^{1\mu} u^2 
         -(\bar{V}_{1\mu}^3-\bar{V}_{2\mu}^3) v_1^{2\mu} u^1
        \right) 
  \nonumber\\
  & & -g_1 \dfrac{f_z}{2} \left(
         \bar{V}_{1\mu}^1 v_1^{2\mu} u_z 
       - \bar{V}_{1\mu}^2 v_1^{1\mu} u_z
       \right).
\label{eq:lag2_uv}
\end{eqnarray}

\subsubsection{Gauge-NGB mixed loop ($uv$)}
We now evaluate the gauge-boson and Nambu-Goldstone-boson mixed loop
diagrams ($uv$ diagrams) in the Landau gauge $\xi=0$.  
From the Lagrangian Eq.(\ref{eq:lag2_uv}), we
obtain the effective action arising from the $uv$ diagrams,
\begin{eqnarray}
  & &
  -\dfrac{3}{2(4\pi)^2 \bar{\epsilon}}
   \left[ g_1^2
         \left((2-\dfrac{f_z^2}{f^2})^2+\dfrac{f_z^2}{f^2}\right)
        + g_2^2 \dfrac{f_z^4}{f^4}
   \right] \dfrac{f^2}{8} 
   ( \bar{V}_{1\mu}^1 \bar{V}_{1}^{1 \mu}
    +\bar{V}_{1\mu}^2 \bar{V}_{1}^{2 \mu} )
  \nonumber\\
  & &
  -\dfrac{3}{(4\pi)^2 \bar{\epsilon}}
    g_1^2 \dfrac{f_z^4}{f^4} \dfrac{f^2}{8} 
    (\bar{V}_{1\mu}^3 - \bar{V}_{2\mu}^3 )
    (\bar{V}_{1}^{3\mu} - \bar{V}_{2}^{3\mu} ),
\end{eqnarray}
which can be simplified to
\begin{eqnarray}
  & & -\dfrac{3}{2(4\pi)^2 \bar{\epsilon}}
      \left[
        g_1^2 (4-3\dfrac{f_z^2}{f^2}+\dfrac{f_z^4}{f^4})
      + g_2^2 \dfrac{f_z^4}{f^4}
      \right] \dfrac{f^2}{4} \tr\left[
        (\bar{V}_{1\mu}-\bar{V}_{2\mu})
        (\bar{V}_{1}^{\mu}-\bar{V}_{2}^{\mu})
      \right]
  \nonumber\\
  & & +\dfrac{3}{4(4\pi)^2 \bar{\epsilon}}
      \left[
        g_1^2 (4-3\dfrac{f_z^2}{f^2}-\dfrac{f_z^4}{f^4})
      + g_2^2 \dfrac{f_z^4}{f^4}
      \right] \dfrac{f^2}{4} 
      \tr\left[(\bar{V}_{1\mu}-\bar{V}_{2\mu}) \tau^3\right]
      \tr\left[(\bar{V}_{1}^{\mu}-\bar{V}_{2}^{\mu}) \tau^3\right] .
  \nonumber\\
  & &
\label{eq:efflag2_uv}
\end{eqnarray}

\subsubsection{NGB loop ($uu$)}
We next consider the Nambu-Goldstone-boson loop effects ($uu$ loops)
generated from the Lagrangian Eq.(\ref{eq:lag2_uu}).
Using the result of Appendix \ref{sec:divergences}, we see that the $uu$ diagrams lead
to one-loop effective action,
\begin{equation}
  \dfrac{1}{(4\pi)^2 \bar{\epsilon}} \left(
    \dfrac{1}{12}\Gamma^{ab}_{\mu\nu} \Gamma^{ba \mu\nu} 
   +\dfrac{1}{2}\sigma^{ab} \sigma^{ba} \right).
\label{eq:efflag2_uu}
\end{equation}
In Landau gauge,  we then obtain
\begin{eqnarray}
\lefteqn{
  \frac{1}{12}\Gamma^{ab}_{\mu\nu}\Gamma^{ba \mu\nu}
        +\frac{1}{2} \sigma^{ab} \sigma^{ba}
} \nonumber\\
  &=& -\dfrac{f_z^2}{12f^2} 
       \tr\left[\bar{V}_{1\mu\nu} \bar{V}_1^{\mu\nu}\right]
      -\dfrac{f_z^4}{12f^4} 
       \tr\left[\bar{V}_{2\mu\nu} \bar{V}_2^{\mu\nu}\right]
  \nonumber\\
  & & -\dfrac{f_z^2}{6 f^2} ( 2 - \dfrac{f_z^2}{f^2} )
       \tr\left[\bar{V}_{1\mu\nu} \bar{V}_2^{\mu\nu}\right]
  \nonumber\\
  & & +\dfrac{f_z^2}{6 f^2} ( 4 - 3\dfrac{f_z^2}{f^2} )
      \, i \,\tr\left[(\bar{V}_{1\mu}-\bar{V}_{2\mu})
            (\bar{V}_{1\nu}-\bar{V}_{2\nu}) \bar{V}_2^{\mu\nu}\right]
  \nonumber\\
  & & +\dfrac{f_z^4}{6 f^4}
     \, i\, \tr\left[(\bar{V}_{1\mu}-\bar{V}_{2\mu})
            (\bar{V}_{1\nu}-\bar{V}_{2\nu}) \bar{V}_1^{\mu\nu}\right]
  \nonumber\\
  & & +\dfrac{1}{6}( 4 - 3\dfrac{f_z^2}{f^2} )^2
      \tr\left[(\bar{V}_{1\mu}-\bar{V}_{2\mu})
            (\bar{V}_{1\nu}-\bar{V}_{2\nu})\right]
      \tr\left[(\bar{V}_{1}^{\mu}-\bar{V}_{2}^{\mu})
            (\bar{V}_{1}^{\nu}-\bar{V}_{2}^{\nu})\right]
  \nonumber\\
  & & +\dfrac{1}{12}(-8 +12\dfrac{f_z^2}{f^2}-3\dfrac{f_z^4}{f^4} )
       \times
  \nonumber\\
  & & \qquad \times
      \tr\left[(\bar{V}_{1\mu}-\bar{V}_{2\mu})
            (\bar{V}_{1}^{\mu}-\bar{V}_{2}^{\mu})\right]
      \tr\left[(\bar{V}_{1\nu}-\bar{V}_{2\nu})
            (\bar{V}_{1}^{\nu}-\bar{V}_{2}^{\nu})\right]
  \nonumber\\
  & & -\dfrac{1}{6}(1-\frac{f_z^2}{f^2})(4-\frac{f_z^2}{f^2})^2
       \times
  \nonumber\\
  & & \qquad \times
      \tr\left[(\bar{V}_{1\mu}-\bar{V}_{2\mu})
          (\bar{V}_{1\nu}-\bar{V}_{2\nu})\right]
      \tr\left[(\bar{V}_{1}^{\mu}-\bar{V}_{2}^{\mu})\tau^3\right]
      \tr\left[(\bar{V}_{1}^{\nu}-\bar{V}_{2}^{\nu})\tau^3\right]
  \nonumber\\
  & &+\dfrac{1}{12}(1-\dfrac{f_z^2}{f^2})
      (8-4\dfrac{f_z^2}{f^2}+5\dfrac{f_z^4}{f^4}) \times
  \nonumber\\
  & & \qquad \times
      \tr\left[(\bar{V}_{1\mu}-\bar{V}_{2\mu})
            (\bar{V}_{1}^{\mu}-\bar{V}_{2}^{\mu})\right]
      \tr\left[(\bar{V}_{1\nu}-\bar{V}_{2\nu})\tau^3\right]
      \tr\left[(\bar{V}_{1}^{\nu}-\bar{V}_{2}^{\nu})\tau^3\right]
  \nonumber\\
  & &-\dfrac{1}{6}(1-\dfrac{f_z^2}{f^2})(4-\dfrac{f_z^2}{f^2})
      \dfrac{1}{4} \tr\left[\bar{V}_{1\mu\nu} \tau^3\right]
                   \tr\left[\bar{V}_{1}^{\mu\nu} \tau^3\right] 
  \nonumber\\
  & &+\dfrac{1}{6}(1-\dfrac{f_z^2}{f^2})(4-\frac{f_z^2}{f^2})
      \,i\, \tr\left[\bar{V}_{1\mu\nu} \tau^3\right]
                  \tr\left[(\bar{V}_{1}^{\mu}-\bar{V}_{2}^{\mu})
                      (\bar{V}_{1}^{\nu}-\bar{V}_{2}^{\nu})\tau^3\right]
  \nonumber\\
  & &+\dfrac{1}{8}(1-\dfrac{f_z^2}{f^2})^2
      (8+4\dfrac{f_z^2}{f^2}+\dfrac{f_z^4}{f^4})
      \dfrac{1}{2}\left[
      \tr\left[(\bar{V}_{1\mu}-\bar{V}_{2\mu})\tau^3\right]
      \tr\left[(\bar{V}_{1}^{\mu}-\bar{V}_{2}^{\mu})\tau^3\right]
      \right]^2 .
\end{eqnarray}

In order to renormalize the one-loop divergences in Eq.(\ref{eq:efflag2_uu}), 
we need to introduce the following counter terms at ${\cal O}(p^4)$,
\begin{eqnarray}
  {\cal L}_{4}
  &=& \alpha_{1} \tr\left[ 
        V_{1\mu\nu} U V_{2}^{\mu\nu} U^\dagger
      \right]
  \nonumber\\
  & & -2i\alpha_{2} \tr\left[
        (D_\mu U)^\dagger (D_\nu U) V_{2\mu\nu}
      \right]
  \nonumber\\
  & & -2i\alpha_{3} \tr\left[
        V_{1}^{\mu\nu} (D_\mu U) (D_\nu U)^\dagger
      \right] 
  \nonumber\\
  & & +\alpha_{4} 
       \tr\left[(D_\mu U) (D_\nu U)^\dagger \right] 
       \tr\left[(D^\mu U) (D^\nu U)^\dagger \right] 
  \nonumber\\
  & & +\alpha_{5} 
       \tr\left[(D_\mu U) (D^\mu U)^\dagger \right] 
       \tr\left[(D_\nu U) (D^\nu U)^\dagger \right]
  \nonumber\\
  & & -\alpha_{6} 
       \tr\left[(D_\mu U) (D_\nu U)^\dagger \right] 
       \tr\left[U^\dagger (D^\mu U) \tau^3 \right] 
       \tr\left[U^\dagger (D^\nu U) \tau^3 \right] 
  \nonumber\\
  & & -\alpha_{7} 
       \tr\left[(D_\mu U) (D^\mu U)^\dagger \right] 
       \tr\left[U^\dagger (D_\nu U) \tau^3 \right] 
       \tr\left[U^\dagger (D^\nu U) \tau^3 \right] 
  \nonumber\\
  & & +\frac{1}{4}\alpha_{8} 
       \tr\left[ U^\dagger V_{1\mu\nu} U \tau^3\right]
       \tr\left[ U^\dagger V^{1\mu\nu} U \tau^3\right]
  \nonumber\\
  & & -\alpha_{9} i
       \tr\left[ U^\dagger V_{1\mu\nu} U \tau^3\right]
       \tr\left[(D^\mu U)^\dagger (D^\nu U) \tau^3 \right] 
  \nonumber\\
  & & +\frac{1}{2} \alpha_{10}
       \left[
       \tr\left[U^\dagger (D_\mu U) \tau^3 \right] 
       \tr\left[U^\dagger (D^\mu U) \tau^3 \right] 
       \right]^2 .
\end{eqnarray}

\subsubsection{$SU(2) \times U(1)$ Renormalization group equations}

We are now ready to perform $\overline{\rm MS}$ renormalization.
For the gauge coupling strengths, we find
\begin{eqnarray}
  \dfrac{1}{g_{1r}^2} 
  &=& \dfrac{1}{g_1^2} -\dfrac{1}{(4\pi)^2 \bar{\epsilon}} \left(
    \dfrac{44}{3} - \dfrac{1}{6}\cdot \dfrac{f^2_z}{f^2} \right),
  \\
  \dfrac{1}{g_{2r}^2} 
  &=& \dfrac{1}{g_2^2} -\dfrac{1}{(4\pi)^2 \bar{\epsilon}} \left(
    - \dfrac{1}{6}\cdot\dfrac{f^4_z}{f^4} \right)~.
\end{eqnarray}
Renormalization of $f$-constants is given by
\begin{eqnarray}
  f_r^2 &=& f^2 
  - \dfrac{3}{2(4\pi)^2 \bar{\epsilon}} f^2
      \left[
        g_1^2 (4-3\dfrac{f_z^2}{f^2}+\dfrac{f_z^4}{f^4})
      + g_2^2 \dfrac{f_z^4}{f^4}
      \right],
  \\
  \beta_r f_r^2 &=& \beta f^2 
  -\dfrac{3}{4(4\pi)^2 \bar{\epsilon}} f^2
      \left[
        g_1^2 (4-3\dfrac{f_z^2}{f^2}-\dfrac{f_z^4}{f^4})
      + g_2^2 \dfrac{f_z^4}{f^4}
      \right] .
\end{eqnarray}
Using Eq.(\ref{eq:f_z}) the above expressions can be simplified as,\footnote{
To leading order in $\beta_r$, the result for the running of $\beta_r$ is consistent with
 that found \protect\cite{Longhitano:1980iz},
in the case of the effective low-energy theory for a standard model with a heavy Higgs
boson.}
\begin{eqnarray}
  f_r^2 &=& f^2 
  -\dfrac{3}{2(4\pi)^2 \bar{\epsilon}} f^2
      \left[
        g_1^2 (2+2\beta+4\beta^2)
      + g_2^2 (1-2\beta)^2
      \right],
  \\
  \beta_r f_r^2 &=& \beta f^2 
  -\dfrac{3}{4(4\pi)^2 \bar{\epsilon}} f^2
      \left[
        g_1^2 (10-4\beta)\beta
      + g_2^2 (1-2\beta)^2
      \right] .
\end{eqnarray}
Finally, the renormalization of ${\cal O}(p^4)$ coefficients $\alpha_i$ is
given by
\begin{equation}
  \alpha_{i}^r = \alpha_i + \dfrac{\gamma_i}{(4\pi)^2 \bar{\epsilon}},
  \qquad
  \mbox{for $i=1,2\cdots, 10$},
\end{equation}
where divergent coefficients $\gamma_i$ are 
\begin{eqnarray}
  \gamma_1 &=& -\dfrac{f_z^2}{6 f^2} ( 2 - \dfrac{f_z^2}{f^2} ) = \dfrac{-1+4\beta^2}{6}, \\
  \gamma_2 &=& -\dfrac{f_z^2}{12 f^2} ( 4 - 3\frac{f_z^2}{f^2} ) = \dfrac{-1-4\beta+12\beta^2}{12},\\
  \gamma_3 &=& -\dfrac{f_z^4}{12 f^4} = \dfrac{-1+4\beta-4\beta^2}{12},\\
  \gamma_4 &=&  \dfrac{1}{6}( 4 - 3\dfrac{f_z^2}{f^2} )^2 =  \dfrac{1+12\beta +36\beta^2}{6},\\
  \gamma_5 &=&  \dfrac{1}{12}(-8 +12\dfrac{f_z^2}{f^2}-3\dfrac{f_z^4}{f^4})=   \dfrac{1-12\beta -12\beta^2}{12}, \\
  \gamma_6 &=& -\dfrac{1}{6}(1-\frac{f_z^2}{f^2})(4-\frac{f_z^2}{f^2})^2  = -\dfrac{\beta(9+12\beta+4\beta^2)}{3},\\
  \gamma_7 &=&  \dfrac{1}{12}(1-\frac{f_z^2}{f^2})
                (8-4\dfrac{f_z^2}{f^2}+5\dfrac{f_z^4}{f^4}) = \dfrac{\beta(9-12\beta+20\beta^2)}{6},\\
  \gamma_8 &=& -\dfrac{1}{6}(1-\dfrac{f_z^2}{f^2})(4-\dfrac{f_z^2}{f^2}) = -\dfrac{\beta(3+2\beta)}{3}, \\
  \gamma_9 &=& -\dfrac{1}{6}(1-\dfrac{f_z^2}{f^2})(4-\dfrac{f_z^2}{f^2}) = -\dfrac{\beta(3+2\beta)}{3},\\
  \gamma_{10} &=&\dfrac{1}{8}(1-\frac{f_z^2}{f^2})^2
                (8+4\dfrac{f_z^2}{f^2}+\frac{f_z^4}{f^4}) = \dfrac{\beta^2(13-12\beta+4\beta^2)}{2}. 
\end{eqnarray}

We then obtain the $\overline{\rm MS}$ renormalization group
equations, 
\begin{eqnarray}
  \mu \dfrac{d}{d\mu} \left(\dfrac{1}{g_{1r}^2}\right)
  &=& \dfrac{1}{(4\pi)^2} 
      \left[ \dfrac{44}{3} - \dfrac{1}{6} (1-2\beta_r)\right],
\label{eq:rge2_g1r}
  \\
  \mu \dfrac{d}{d\mu} \left(\dfrac{1}{g_{2r}^2}\right)
  &=& \dfrac{1}{(4\pi)^2} 
      \left[ - \dfrac{1}{6}(1-2\beta_r)^2 \right],
\label{eq:rge2_g2r}
\end{eqnarray}
for the gauge coupling strengths, and 
\begin{eqnarray}
  \mu \dfrac{d}{d\mu} f_r^2 
  &=& \dfrac{3}{2(4\pi)^2} \left[
        g_{1r}^2 (2+2\beta_r+4\beta_r^2)
      + g_{2r}^2 (1-2\beta_r)^2
      \right]  f_r^2, 
\label{eq:rge2_f}
  \\
  \mu \dfrac{d}{d\mu} (\beta_r f_r^2) 
  &=& \dfrac{3}{4(4\pi)^2} \left[
        g_{1r}^2 (10-4\beta_r)\beta_r
      + g_{2r}^2 (1-2\beta_r)^2
      \right] f_r^2,
\label{eq:rge2_beta}
\end{eqnarray}
for the $f$-constants.
It should be emphasized that, even if we start with
$\beta_r(\mu=\Lambda)=0$ at the cutoff scale,  non-vanishing $\beta_r$
is generated at lower energies through the $g_{2r}^2$ term in the renormalization group 
Eq.(\ref{eq:rge2_beta}).
We also note that the $U(1)_2$ gauge field decouples from the
non-linear sigma model in the $\beta_r = 1/2$ limit.  The $g_{2r}^2$
terms thus vanish in Eqs.(\ref{eq:rge2_f}) and (\ref{eq:rge2_beta}) 
in the $\beta_r = 1/2$ limit.
Actually, we can show that $\beta_r=1/2$ is a fixed point in the
renormalization group Eqs.(\ref{eq:rge2_f}) and (\ref{eq:rge2_beta}).

The renormalization group equations for 
the ${\cal O}(p^4)$ parameters is given by
\begin{eqnarray}
  \mu \dfrac{d}{d\mu} \alpha_1^r
  &=& \dfrac{1-4\beta_r^2}{6(4\pi)^2}, 
\label{eq:rge2_alpha1}
  \\
  \mu \dfrac{d}{d\mu} \alpha_2^r
  &=& \dfrac{1+4\beta_r-12\beta_r^2}{12(4\pi)^2},
\label{eq:rge2_alpha2}
  \\
  \mu \dfrac{d}{d\mu} \alpha_3^r
  &=& \dfrac{1-4\beta_r+4\beta_r^2}{12(4\pi)^2},
\label{eq:rge2_alpha3}
  \\
  \mu \dfrac{d}{d\mu} \alpha_4^r
  &=& \dfrac{-1-12\beta_r -36\beta_r^2}{6(4\pi)^2},
\label{eq:rge2_alpha4}
  \\
  \mu \dfrac{d}{d\mu} \alpha_5^r
  &=& \dfrac{-1+12\beta_r +12\beta_r^2}{12(4\pi)^2},
\label{eq:rge2_alpha5}
  \\
  \mu \dfrac{d}{d\mu} \alpha_6^r
  &=& \dfrac{\beta_r(9+12\beta_r+4\beta_r^2)}{3(4\pi)^2},
\label{eq:rge2_alpha6}
  \\
  \mu \dfrac{d}{d\mu} \alpha_7^r
  &=& \dfrac{\beta_r(-9+12\beta_r-20\beta_r^2)}{6(4\pi)^2},
\label{eq:rge2_alpha7}
  \\
  \mu \dfrac{d}{d\mu} \alpha_8^r
  &=& \dfrac{\beta_r(3+2\beta_r)}{3(4\pi)^2},
\label{eq:rge2_alpha8}
  \\
  \mu \dfrac{d}{d\mu} \alpha_9^r
  &=& \dfrac{\beta_r(3+2\beta_r)}{3(4\pi)^2},
\label{eq:rge2_alpha9}
  \\
  \mu \dfrac{d}{d\mu} \alpha_{10}^r
  &=& \dfrac{\beta_r^2(-13+12\beta_r-4\beta_r^2)}{2(4\pi)^2} .
\label{eq:rge2_alpha10}
\end{eqnarray}
The Eqs.(\ref{eq:rge2_alpha1})--(\ref{eq:rge2_alpha10}) agree with 
the $SU(2) \times SU(2)$ chiral perturbation theory results in the $\beta_r=0$ limit.
We also note that Eqs.(\ref{eq:rge2_alpha6})--(\ref{eq:rge2_alpha9})
are proportional to $\beta_r$ for $\beta_r \ll 1$, while
Eq.(\ref{eq:rge2_alpha10}) is proportional to $\beta_r^2$.
These behaviors are consistent with that expected from  custodial symmetry.

The RGE equations in the two site regime, 
Eqs.~(\ref{eq:twositebegin})-(\ref{eq:twositeend}), follow immediately
in the small $\beta_r$ limit.

\subsection{The Three Site Model}
\label{sec:threesiteappendix}

Finally, we discuss the $SU(2)\times SU(2)\times U(1)$ three site moose
model shown in Figure~\ref{fig:threesite}.
To lowest order (${\cal O}(p^2)$) Lagrangian is given by
\begin{eqnarray}
  {\cal L}_2 &=& 
    \dfrac{f_1^2}{4} 
    \tr\left[(D_\mu U_1)^\dagger (D^\mu U_1) \right]
   +\dfrac{f_2^2}{4} 
    \tr\left[(D_\mu U_2)^\dagger (D^\mu U_2) \right]
  \nonumber\\
  & &
  +\beta_{(2)}\dfrac{f_2^2}{4} 
      \tr\left[  U_2^\dagger (D_\mu U_2) \tau_3 \right]
      \tr\left[  U_2^\dagger (D^\mu U_2) \tau_3 \right]
  \nonumber\\
  & &
   -\dfrac{1}{2g_0^2}
    \tr\left[V_{0\mu\nu}V_0^{\mu\nu} \right]
   -\dfrac{1}{2g_1^2}
    \tr\left[V_{1\mu\nu}V_1^{\mu\nu} \right]
   -\dfrac{1}{2g_2^2}
    \tr\left[V_{2\mu\nu}V_2^{\mu\nu} \right],
\label{eq:221p2}
\end{eqnarray}
with $U_1$, $U_2$ being chiral fields (at link-1 and link-2 in the
moose diagram Figure~\ref{fig:threesite}),
\begin{equation}
  U_1 = \exp\left[\dfrac{2 i \pi_1 T^a}{f} \right], \qquad
  U_2 = \exp\left[\dfrac{2 i \pi_2 T^a}{f} \right].
\end{equation}
The covariant derivatives are given by
\begin{equation}
  D_\mu U_1 = \partial_\mu U_1 + iV_{0\mu} U_1 - i U_1 V_{1\mu}, 
  \qquad
  D_\mu U_2 = \partial_\mu U_2 + iV_{1\mu} U_2 -i  U_2 V_{2\mu}, 
\end{equation}
where
$V_{0\mu}$, $V_{1\mu}$ and $V_{2\mu}$ are gauge fields at site-0,
site-1 and site-2,
\begin{equation}
  V_{0\mu} \equiv V^a_{0\mu} T^a, \qquad
  V_{1\mu} \equiv V^a_{1\mu} T^a, \qquad
  V_{2\mu} \equiv V^3_{2\mu} T^3.
\end{equation}
Note that $V_{0\mu}$ and $V_{1\mu}$ are $SU(2)$ gauge fields, while 
$V_{2\mu}$ belong to $U(1)$.

In the previous sections, we have investigated the one-loop logarithmic divergences
which appear in $SU(2) \times SU(2)$ and in $SU(2) \times U(1)$ gauged nonlinear
sigma models. In essence, therefore, we have investigated the divergences associated
with each of the links in the three site model {\it separately}. Naively, one may expect to
proceed in the three site model by simply adding the divergences associated with the 
these two links together. This, in fact,
turns out to be the case. For this to be the case, however, we must show
that there are no {\it new} divergences at one-loop which are intrinsic to
the three site model -- in particular, that there are no
 {\em next-to-nearest-neighbor}  (NNN) operators, for
 example the ${\cal  O}(p^2)$ term
\begin{equation}
  \tr\left[U_1^\dagger (D_\mu U_1) (D^\mu U_2) U_2^\dagger\right]~,
\label{eq:nnn_int}
\end{equation}
which are induced at one-loop from the ${\cal O}(p^2)$ interactions
included in the three site model.

Consider an NNN term, such as that in Eq.(\ref{eq:nnn_int}). By
definition, such a term contains both chiral
fields at link-1 and at link-2. Note that the three site model separates
into two decoupled models in the limit that $g^2_1 \to 0$. 
Therefore  we see that any  NNN term would be generated at one-loop
proportional to $g_1^2$ and, therefore, 
in  Landau gauge (in which pions are massless) this means that
NNN terms can only  be generated by a
$V_{1\mu}$ gauge boson loop. From the form of the three site
Lagrangian, Eq.(\ref{eq:221p2}), we see that there are no
 $V_{1\mu}V_1^{\mu} (\pi_1)^n$ nor
$V_{1\mu}V_1^{\mu} (\pi_2)^n$ interactions ($n \ge 1$) in our
three site Lagrangian. 
Therefore a $v_1v_1$-loop diagram cannot produce an NNN term.
In addition, we see that the first term in Eq.(\ref{eq:221p2}) contains $u_1 v_1$
interactions, while the second term in Eq.(\ref{eq:221p2}) has $u_2 v_1$
interactions.
In the Landau gauge, however, there is no $u_1$-$u_2$ mixing.
Therefore $u v_1$-loop diagrams cannot produce an
NNN term.
The remaining possibilities are $uv_0$ (and $u v_2$) diagrams, in
which the $v_0$ (or $v_2$) mix with $v_1$.
The integrand of such a one-loop diagram, however,  will be highly suppressed in the
ultraviolet  and will not produce logarithmic divergences.
We thus conclude that NNN terms
are not generated through one-loop
diagrams.\footnote{This observation is consistent with the results of HLS loop
calculations \protect\cite{Harada:2003jx}, in which it is found that Georgi's vector-limit ($a=1$)
-- which corresponds to the  three site linear moose model --  is a 
renormalization group fixed point of the mass independent
renormalization group equations.}

\subsubsection{Renormalization group equations}

We may now derive the renormalization group equations in the
three site model by combining the results presented in the previous sections. 
We then immediately find
\begin{eqnarray}
  \mu \dfrac{d}{d\mu} \left(\dfrac{1}{g_{0r}^2}\right)
  &=& \dfrac{1}{(4\pi)^2} 
      \left[ \dfrac{44}{3} - \dfrac{1}{6} \right],
\label{eq:rge3_g0r}
  \\
  \mu \dfrac{d}{d\mu} \left(\dfrac{1}{g_{1r}^2}\right)
  &=& \dfrac{1}{(4\pi)^2} 
      \left[ \dfrac{44}{3} - \dfrac{1}{6} - \dfrac{1}{6}(1-2\beta_{(2)r}) \right],
\label{eq:rge3_g1r}
  \\
  \mu \dfrac{d}{d\mu} \left(\dfrac{1}{g_{2r}^2}\right)
  &=& \dfrac{1}{(4\pi)^2} 
      \left[ - \dfrac{1}{6} (1-2\beta_{(2)r})^2\right],
\label{eq:rge3_g2r}
\end{eqnarray}
for the renormalized gauge coupling strengths.
Here the first $-1/6$ in Eq.(\ref{eq:rge3_g1r}) comes from a $u_1 u_1$-loop,
while the second $-1/6$ comes from  a $u_2 u_2$ loop.
For the renormalized $f$-constants, we find
\begin{eqnarray}
  \mu \dfrac{d}{d\mu} f_{1r}^2 
    &=& \dfrac{3}{(4\pi)^2} (g_{0r}^2 + g_{1r}^2) f_{1r}^2, 
    \\
\label{eq:rge3_f1}
  \mu \dfrac{d}{d\mu} f_{2r}^2 
  &=& \dfrac{3}{2(4\pi)^2} \left[
        g_{1r}^2 (2+2\beta_{(2)r}+4\beta_{(2)r}^2)
      + g_{2r}^2 (1-2\beta_{(2)r})^2
      \right]  f_{2r}^2, 
\label{eq:rge3_f2}
  \\
  \mu \dfrac{d}{d\mu} (\beta_{(2)r} f_{2r}^2) 
  &=& \dfrac{3}{4(4\pi)^2} \left[
        g_{1r}^2 (10-4\beta_{(2)r})\beta_{(2)r}
      + g_{2r}^2 (1-2\beta_{(2)r})^2
      \right] f_{2r}^2.
\label{eq:rge3_beta}
\end{eqnarray}

We need to introduce the ${\cal O}(p^4)$ counter terms,
\begin{eqnarray}
  {\cal L}_{(1)4}
  &=& \alpha_{(1)1} \tr\left[ 
        V_{0\mu\nu} U_1 V_{1}^{\mu\nu} U_1^\dagger
      \right]
  \nonumber\\
  & & -2i\alpha_{(1)2} \tr\left[
        (D_\mu U_1)^\dagger (D_\nu U_1) V_{1\mu\nu}
      \right]
  \nonumber\\
  & & -2i\alpha_{(1)3} \tr\left[
        V_{0}^{\mu\nu} (D_\mu U_1) (D_\nu U_1)^\dagger
      \right] 
  \nonumber\\
  & & +\alpha_{(1)4} 
       \tr\left[(D_\mu U_1) (D_\nu U_1)^\dagger \right] 
       \tr\left[(D^\mu U_1) (D^\nu U_1)^\dagger \right] 
  \nonumber\\
  & & +\alpha_{(1)5} 
       \tr\left[(D_\mu U_1) (D^\mu U_1)^\dagger \right] 
       \tr\left[(D_\nu U_1) (D^\nu U_1)^\dagger \right],
\end{eqnarray}
and
\begin{eqnarray}
  {\cal L}_{(2)4}
  &=& \alpha_{(2)1} \tr\left[ 
        V_{1\mu\nu} U_2 V_{2}^{\mu\nu} U_2^\dagger
      \right]
  \nonumber\\
  & & -2i\alpha_{(2)2} \tr\left[
        (D_\mu U_2)^\dagger (D_\nu U_2) V_{2\mu\nu}
      \right]
  \nonumber\\
  & & -2i\alpha_{(2)3} \tr\left[
        V_{1}^{\mu\nu} (D_\mu U_2) (D_\nu U_2)^\dagger
      \right] 
  \nonumber\\
  & & +\alpha_{(2)4} 
       \tr\left[(D_\mu U_2) (D_\nu U_2)^\dagger \right] 
       \tr\left[(D^\mu U_2) (D^\nu U_2)^\dagger \right] 
  \nonumber\\
  & & +\alpha_{(2)5} 
       \tr\left[(D_\mu U_2) (D^\mu U_2)^\dagger \right] 
       \tr\left[(D_\nu U_2) (D^\nu U_2)^\dagger \right]
  \nonumber\\
  & & -\alpha_{(2)6} 
       \tr\left[(D_\mu U_2) (D_\nu U_2)^\dagger \right] 
       \tr\left[U_2^\dagger (D^\mu U_2) \tau^3 \right] 
       \tr\left[U_2^\dagger (D^\nu U_2) \tau^3 \right] 
  \nonumber\\
  & & -\alpha_{(2)7} 
       \tr\left[(D_\mu U_2) (D^\mu U_2)^\dagger \right] 
       \tr\left[U_2^\dagger (D_\nu U_2) \tau^3 \right] 
       \tr\left[U_2^\dagger (D^\nu U_2) \tau^3 \right] 
  \nonumber\\
  & & +\frac{1}{4}\alpha_{(2)8} 
       \tr\left[ U_2^\dagger V_{1\mu\nu} U_2 \tau^3\right]
       \tr\left[ U_2^\dagger V^{1\mu\nu} U_2 \tau^3\right]
  \nonumber\\
  & & -\alpha_{(2)9} i
       \tr\left[ U_2^\dagger V_{1\mu\nu} U_2 \tau^3\right]
       \tr\left[(D^\mu U_2)^\dagger (D^\nu U_2) \tau^3 \right] 
  \nonumber\\
  & & +\frac{1}{2} \alpha_{(2)10}
       \left[
       \tr\left[U_2^\dagger (D_\mu U_2) \tau^3 \right] 
       \tr\left[U_2^\dagger (D^\mu U_2) \tau^3 \right] 
       \right]^2 .
\end{eqnarray}
The renormalization group equations for these
coefficients are 
\begin{eqnarray}
  \mu \dfrac{d}{d\mu} \alpha_{(1)1}^r
  &=& \dfrac{1}{6(4\pi)^2}, 
  \\
  \mu \dfrac{d}{d\mu} \alpha_{(1)2}^r 
  &=& \dfrac{1}{12(4\pi)^2}, 
  \\
  \mu \dfrac{d}{d\mu} \alpha_{(1)3}^r 
  &=& \dfrac{1}{12(4\pi)^2}, 
  \\
  \mu \dfrac{d}{d\mu} \alpha_{(1)4}^r 
  &=& -\dfrac{1}{6(4\pi)^2}, 
  \\
  \mu \dfrac{d}{d\mu} \alpha_{(1)5}^r 
  &=& -\dfrac{1}{12(4\pi)^2}, 
\end{eqnarray}
and
\begin{eqnarray}
  \mu \dfrac{d}{d\mu} \alpha_{(2)1}^r
  &=& \dfrac{1-4\beta_{(2)r}^2}{6(4\pi)^2}, 
  \\
  \mu \dfrac{d}{d\mu} \alpha_{(2)2}^r
  &=& \dfrac{1+4\beta_{(2)r}-12\beta_{(2)r}^2}{12(4\pi)^2},
  \\
  \mu \dfrac{d}{d\mu} \alpha_{(2)3}^r
  &=& \dfrac{1-4\beta_{(2)r}+4\beta_{(2)r}^2}{12(4\pi)^2},
  \\
  \mu \dfrac{d}{d\mu} \alpha_{(2)4}^r
  &=& \dfrac{-1-12\beta_{(2)r} -36\beta_{(2)r}^2}{6(4\pi)^2},
  \\
  \mu \dfrac{d}{d\mu} \alpha_{(2)5}^r
  &=& \dfrac{-1+12\beta_{(2)r} +12\beta_{(2)r}^2}{12(4\pi)^2},
  \\
  \mu \dfrac{d}{d\mu} \alpha_{(2)6}^r
  &=& \dfrac{\beta_{(2)r}(9+12\beta_{(2)r}+4\beta_{(2)r}^2)}{3(4\pi)^2},
  \\
  \mu \dfrac{d}{d\mu} \alpha_{(2)7}^r
  &=& \dfrac{\beta_{(2)r}(-9+12\beta_{(2)r}-20\beta_{(2)r}^2)}{6(4\pi)^2},
  \\
  \mu \dfrac{d}{d\mu} \alpha_{(2)8}^r
  &=& \dfrac{\beta_{(2)r}(3+2\beta_{(2)r})}{3(4\pi)^2},
  \\
  \mu \dfrac{d}{d\mu} \alpha_{(2)9}^r
  &=& \dfrac{\beta_{(2)r}(3+2\beta_{(2)r})}{3(4\pi)^2},
  \\
  \mu \dfrac{d}{d\mu} \alpha_{(2)10}^r
  &=& \dfrac{\beta_{(2)r}^2(-13+12\beta_{(2)r}-4\beta_{(2)r}^2)}{2(4\pi)^2} .
\end{eqnarray}

Finally, we  consider the renormalization group property of the
delocalization operator, which only depends on the first link, and
we find
\begin{equation}
  x_1 \tr\left[ J^\mu D_\mu U_1 U_1^\dagger \right].
\end{equation}
We find
\begin{equation}
  \mu \dfrac{d}{d\mu} x_1^r = \dfrac{3g_{1r}^2}{(4\pi)^2} x_1^r.
\end{equation}

The three site RGE equations used in the body of the paper, Eqs.~(\ref{eq:threesitefbegin})-(\ref{eq:threesitedelocal}), follow immediately in the small $\beta_r$ limit. The three site RGE
equations for $\alpha_{(i)1-5}$ arise solely from NGB loops, and are therefore identical
with those calculated \cite{Tanabashi:1993sr} 
in hidden local symmetry \cite{Bando:1985ej,Bando:1985rf,Bando:1988ym,Bando:1988br,Harada:2003jx} models of QCD in the ``vector limit" \cite{Georgi:1989xy}.

\section{Divergences in scalar field one-loop integrals}
\label{sec:divergences}

In this appendix, we consider one-loop integrals of the scalar field
$u$,
\begin{equation}
  {\cal L} = \frac{1}{2} (D_\mu^{ab} u^b) (D_\mu^{ac} u^c)
            -\frac{1}{2} \sigma^{ac} u^a u^c,
\end{equation}
where covariant derivative is given by
\begin{equation}
  D_\mu^{ab} u^b = \partial_\mu u^a + \Gamma^{ab}_\mu u^b.
\end{equation}
We also assume
\begin{equation}
  \sigma^{ab} = \sigma^{ba}.
\end{equation}
The logarithmic divergence in the one-loop  effective
Lagrangian is calculated to be
\begin{equation}
  \dfrac{1}{(4\pi)^2 \bar{\epsilon}} \left(
    \frac{1}{12} \Gamma_{\mu\nu}^{ab} \Gamma^{ba\mu\nu}
   +\frac{1}{2} \sigma^{ab}\sigma^{ba}
    \right),
\end{equation}
with $\Gamma^{ab}_{\mu\nu}$ being given by
\begin{equation}
  \Gamma^{ab}_{\mu\nu} 
    = \partial_\mu \Gamma_\nu^{ab}-\partial_\nu \Gamma_\mu^{ab}
     +[\Gamma_\mu, \Gamma_\nu]^{ab}.
\end{equation}
Here $\bar{\epsilon}$ is defined as
\begin{equation}
  \dfrac{1}{\bar{\epsilon}} \equiv 
  \dfrac{\Gamma(2-d/2)}{2(4\pi)^{d/2-2}},
\end{equation}
with $d$ being the dimensionality of space-time.



\begin{thebibliography}{99}


\bibitem{Csaki:2003dt}
  C.~Csaki, C.~Grojean, H.~Murayama, L.~Pilo and J.~Terning,
{\it Gauge theories on an interval: Unitarity without a Higgs},
  Phys.\ Rev.\ D {\bf 69}, 055006 (2004)
  [arXiv:hep-ph/0305237].

\bibitem{Higgs:1964ia}
P.~W. Higgs, {\it Broken symmetries, massless particles and gauge fields},
  {\em Phys. Lett.} {\bf 12} (1964) 132--133.

 \bibitem{SekharChivukula:2001hz}
R.~Sekhar~Chivukula, D.~A. Dicus, and H.-J. He, {\it Unitarity of compactified
  five dimensional yang-mills theory},  {\em Phys. Lett.} {\bf B525} (2002)
  175--182, [arXiv:hep-ph/0111016].

\bibitem{Chivukula:2002ej}
R.~S. Chivukula and H.-J. He, {\it Unitarity of deconstructed five-dimensional
  yang-mills theory},  {\em Phys. Lett.} {\bf B532} (2002) 121--128,
  [arXiv:hep-ph/0201164].

\bibitem{Chivukula:2003kq}
R.~S. Chivukula, D.~A. Dicus, H.-J. He, and S.~Nandi, {\it Unitarity of the
  higher dimensional standard model},  {\em Phys. Lett.} {\bf B562} (2003)
  109--117, [arXiv:hep-ph/0302263].

\bibitem{He:2004zr}
H.-J.~He,
{\it Higgsless deconstruction without boundary condition},
arXiv:hep-ph/0412113.

\bibitem{Antoniadis:1990ew}
  I.~Antoniadis,
  Phys.\ Lett.\ B {\bf 246}, 377 (1990).



%
\bibitem{Agashe:2003zs}
  K.~Agashe, A.~Delgado, M.~J.~May and R.~Sundrum,
  {\it RS1, Custodial Isospin and Precision Tests},
  JHEP {\bf 0308}, 050 (2003)
  [arXiv:hep-ph/0308036].

\bibitem{Csaki:2003zu}
C.~Csaki, C.~Grojean, L.~Pilo, and J.~Terning, {\it Towards a realistic model
  of higgsless electroweak symmetry breaking},  {\em Phys. Rev. Lett.} {\bf 92}
  (2004) 101802, [arXiv:hep-ph/0308038].

\bibitem{Burdman:2003ya}
  G.~Burdman and Y.~Nomura,
 {\it Holographic theories of electroweak symmetry breaking without a Higgs
  boson},
  Phys.\ Rev.\ D {\bf 69}, 115013 (2004)
  [arXiv:hep-ph/0312247].

\bibitem{Cacciapaglia:2004jz}
  G.~Cacciapaglia, C.~Csaki, C.~Grojean and J.~Terning,
  {\it Oblique corrections from Higgsless models in warped space},
  {\em Phys. Rev. D} {\bf 70}, (2004) 075014,
  [arXiv:hep-ph/0401160].

\bibitem{Arkani-Hamed:2001ca}
N.~Arkani-Hamed, A.~G. Cohen, and H.~Georgi, {\it (de)constructing dimensions},
   {\em Phys. Rev. Lett.} {\bf 86} (2001) 4757--4761,
  [arXiv:hep-th/0104005].

\bibitem{Hill:2000mu}
C.~T. Hill, S.~Pokorski, and J.~Wang, {\it Gauge invariant effective lagrangian
  for kaluza-klein modes},  {\em Phys. Rev.} {\bf D64} (2001) 105005,
  [arXiv:hep-th/0104035].


 \bibitem{Foadi:2003xa}
R.~Foadi, S.~Gopalakrishna, and C.~Schmidt, {\it Higgsless electroweak symmetry
  breaking from theory space},  {\em JHEP} {\bf 03} (2004) 042,
  [arXiv: hep-ph/0312324].

\bibitem{Hirn:2004ze}
  J.~Hirn and J.~Stern,
  {\it The role of spurions in Higgs-less electroweak effective theories},
  Eur.\ Phys.\ J.\ C {\bf 34}, 447 (2004)
  [arXiv:hep-ph/0401032].

\bibitem{Casalbuoni:2004id}
R.~Casalbuoni, S.~De Curtis and D.~Dominici,
{\it Moose models with vanishing S parameter},
Phys.\ Rev.\ D {\bf 70} (2004) 055010
[arXiv:hep-ph/0405188].

\bibitem{Chivukula:2004pk}
R.~S.~Chivukula, E.~H.~Simmons, H.~J.~He, M.~Kurachi and M.~Tanabashi,
{\it The structure of corrections to electroweak interactions in Higgsless
models},
Phys.\ Rev.\ D {\bf 70} (2004) 075008
[arXiv:hep-ph/0406077].


\bibitem{Perelstein:2004sc}
M.~Perelstein, {\it Gauge-assisted technicolor?},  {\em JHEP} {\bf 10} (2004)
  010, [arXiv:hep-ph/0408072].

\bibitem{Georgi:2004iy}
  H.~Georgi, {\it Fun with Higgsless theories},
  Phys.\ Rev.\ D {\bf 71}, 015016 (2005)
  [arXiv:hep-ph/0408067].


\bibitem{SekharChivukula:2004mu}
R.~Sekhar Chivukula, E.~H.~Simmons, H.~J.~He, M.~Kurachi and M.~Tanabashi,
{\it Electroweak corrections and unitarity in linear moose models},
Phys.\ Rev.\ D {\bf 71} (2005) 035007
[arXiv:hep-ph/0410154].

\bibitem{SekharChivukula:2006cg}
  R.~Sekhar Chivukula, B.~Coleppa, S.~Di Chiara, E.~H.~Simmons, H.~J.~He, M.~Kurachi and M.~Tanabashi,
 {\it A three site higgsless model},
  Phys.\ Rev.\ D {\bf 74}, 075011 (2006)
  [arXiv:hep-ph/0607124].





  \bibitem{Casalbuoni:1985kq}
R.~Casalbuoni, S.~De~Curtis, D.~Dominici, and R.~Gatto, {\it Effective weak
  interaction theory with possible new vector resonance from a strong higgs
  sector},  {\em Phys. Lett.} {\bf B155} (1985) 95.

\bibitem{Casalbuoni:1996qt}
R.~Casalbuoni {\em et.~al.}, {\it Degenerate bess model: The possibility of a
  low energy strong electroweak sector},  {\em Phys. Rev.} {\bf D53} (1996)
  5201--5221, [\href{http://xxx.lanl.gov/abs/hep-ph/9510431}{{\tt
  hep-ph/9510431}}].

  \bibitem{Bando:1985ej}
M.~Bando, T.~Kugo, S.~Uehara, K.~Yamawaki, and T.~Yanagida, {\it Is rho meson a
  dynamical gauge boson of hidden local symmetry?},  {\em Phys. Rev. Lett.}
  {\bf 54} (1985) 1215.

\bibitem{Bando:1985rf}
M.~Bando, T.~Kugo, and K.~Yamawaki, {\it On the vector mesons as dynamical
  gauge bosons of hidden local symmetries},  {\em Nucl. Phys.} {\bf B259}
  (1985) 493.

\bibitem{Bando:1988ym}
M.~Bando, T.~Fujiwara, and K.~Yamawaki, {\it Generalized hidden local symmetry
  and the a1 meson},  {\em Prog. Theor. Phys.} {\bf 79} (1988) 1140.

\bibitem{Bando:1988br}
M.~Bando, T.~Kugo, and K.~Yamawaki, {\it Nonlinear realization and hidden local
  symmetries},  {\em Phys. Rept.} {\bf 164} (1988) 217--314.

\bibitem{Harada:2003jx}
M.~Harada and K.~Yamawaki, {\it Hidden local symmetry at loop: A new
  perspective of composite gauge boson and chiral phase transition},  {\em
  Phys. Rept.} {\bf 381} (2003) 1--233,
  [\href{http://xxx.lanl.gov/abs/hep-ph/0302103}{{\tt hep-ph/0302103}}].


\bibitem{Cacciapaglia:2004rb}
G.~Cacciapaglia, C.~Csaki, C.~Grojean and J.~Terning,
{\it Curing the ills of Higgsless models: The S parameter and unitarity},
Phys.\ Rev.\ D {\bf 71} (2005) 035015
[arXiv:hep-ph/0409126].

\bibitem{Cacciapaglia:2005pa}
  G.~Cacciapaglia, C.~Csaki, C.~Grojean, M.~Reece and J.~Terning,
 {\it Top and bottom: A brane of their own},
  {\em Phys.  Rev.  D}  {\bf 72}, (2005) 095018
  [arXiv:hep-ph/0505001].

\bibitem{Foadi:2004ps}
R.~Foadi, S.~Gopalakrishna and C.~Schmidt,
{\it Effects of fermion localization in Higgsless theories and electroweak
constraints},
Phys.\ Lett.\ B {\bf 606} (2005) 157
[arXiv:hep-ph/0409266].

\bibitem{Foadi:2005hz}
  R.~Foadi and C.~Schmidt,
 {\it An Effective Higgsless Theory: Satisfying Electroweak Constraints and a
  Heavy Top Quark},
  Phys.\ Rev.\ D {\bf 73} (2006)  075011
  [arXiv:hep-ph/0509071].

\bibitem{Chivukula:2005bn}
  R.~S.~Chivukula, E.~H.~Simmons, H.~J.~He, M.~Kurachi and M.~Tanabashi,
   {\it Deconstructed Higgsless models with one-site delocalization},
  %
  Phys.\ Rev.\ D {\bf 71}, 115001 (2005)
  [arXiv:hep-ph/0502162].

\bibitem{Casalbuoni:2005rs}
R.~Casalbuoni, S.~De Curtis, D.~Dolce and D.~Dominici,
{\it Playing with fermion couplings in Higgsless models},
Phys.\ Rev.\ D {\bf 71}, 075015 (2005)
[arXiv:hep-ph/0502209].


\bibitem{SekharChivukula:2005xm}
  R.~Sekhar Chivukula, E.~H.~Simmons, H.~J.~He, M.~Kurachi and M.~Tanabashi,
  {\it Ideal fermion delocalization in Higgsless models},
    Phys.\ Rev.\ D {\bf 72}, 015008 (2005)
  [arXiv:hep-ph/0504114].




\bibitem{Peskin:1992sw}
M.~E. Peskin and T.~Takeuchi, {\it Estimation of oblique electroweak
  corrections},  {\em Phys. Rev.} {\bf D46} (1992) 381--409.

\bibitem{Altarelli:1990zd}
G.~Altarelli and R.~Barbieri, {\it Vacuum polarization effects of new physics
  on electroweak processes},  {\em Phys. Lett.} {\bf B253} (1991) 161--167.

\bibitem{Altarelli:1991fk}
G.~Altarelli, R.~Barbieri, and S.~Jadach, {\it Toward a model independent
  analysis of electroweak data},  {\em Nucl. Phys.} {\bf B369} (1992) 3--32.

\bibitem{Barbieri:2004qk}
  R.~Barbieri, A.~Pomarol, R.~Rattazzi and A.~Strumia,
 {\it Electroweak symmetry breaking after LEP1 and LEP2},
  Nucl.\ Phys.\ B {\bf 703}, 127 (2004)
  [arXiv:hep-ph/0405040].
  

  \bibitem{Chivukula:2004af}
R.~S. Chivukula, E.~H. Simmons, H.-J. He, M.~Kurachi, and M.~Tanabashi, {\it
  Universal non-oblique corrections in higgsless models and beyond},  {\em
  Phys. Lett.} {\bf B603} (2004) 210--218,
  [arXiv:hep-ph/0408262].


\bibitem{Matsuzaki:2006wn}
  S.~Matsuzaki, R.~S.~Chivukula, E.~H.~Simmons and M.~Tanabashi,
  {\it One-Loop Corrections to the S and T Parameters in a Three Site Higgsless
  Model},
  arXiv:hep-ph/0607191.


\bibitem{Degrassi:1992ue}
  G.~Degrassi and A.~Sirlin,
  Phys.\ Rev.\ D {\bf 46}, 3104 (1992).

\bibitem{Degrassi:1992ff}
  G.~Degrassi and A.~Sirlin,
  Nucl.\ Phys.\ B {\bf 383}, 73 (1992). 

\bibitem{Tomohiro}
T.~Abe and M.~Tanabashi, private communication.


\bibitem{Appelquist:1980ae}
  T.~Appelquist and C.~W.~Bernard,
   {\it The Nonlinear Sigma Model In The Loop Expansion},
  %
  Phys.\ Rev.\ D {\bf 23}, 425 (1981).


\bibitem{Appelquist:1980vg}
  T.~Appelquist and C.~W.~Bernard,
 {\it Strongly Interacting Higgs Bosons},
  Phys.\ Rev.\ D {\bf 22}, 200 (1980).

\bibitem{Longhitano:1980iz}
  A.~C.~Longhitano,
  Phys.\ Rev.\  D {\bf 22}, 1166 (1980).
  
\bibitem{Longhitano:1980tm}
  A.~C.~Longhitano,
  {\it Low-Energy Impact Of A Heavy Higgs Boson Sector},
  Nucl.\ Phys.\  B {\bf 188}, 118 (1981).

\bibitem{Appelquist:1993ka}
  T.~Appelquist and G.~H.~Wu,
 {\it The Electroweak chiral Lagrangian and new precision
  measurements},
  Phys.\ Rev.\  D {\bf 48}, 3235 (1993)
  [arXiv:hep-ph/9304240].

\bibitem{Weinberg:1978kz}
  S.~Weinberg,
  Physica A {\bf 96} (1979) 327.


\bibitem{Gasser:1983yg}
  J.~Gasser and H.~Leutwyler,
  Annals Phys.\  {\bf 158}, 142 (1984).

\bibitem{Gasser:1984gg}
  J.~Gasser and H.~Leutwyler,
  Nucl.\ Phys.\  B {\bf 250}, 465 (1985).

\bibitem{Herrero:1993nc}
  M.~J.~Herrero and E.~Ruiz Morales,
  Nucl.\ Phys.\  B {\bf 418}, 431 (1994)
  [arXiv:hep-ph/9308276].

\bibitem{Dobado:1997jx}
  A.~Dobado, A.~Gomez-Nicola, A.~Maroto and J.~R.~Pelaez,
{\it  N.Y., Springer-Verlag, 1997. (Texts and Monographs in Physics)}



\bibitem{Georgi:1985hf}
H.~Georgi, {\it A tool kit for builders of composite models},  {\em Nucl.
  Phys.} {\bf B266} (1986) 274.

\bibitem{Anichini:1994xx}
  L.~Anichini, R.~Casalbuoni and S.~De Curtis,
  Phys.\ Lett.\ B {\bf 348}, 521 (1995)
  [arXiv:hep-ph/9410377].


\bibitem{Maldacena:1998re}
J.~M. Maldacena, {\it The large n limit of superconformal field theories and
  supergravity},  {\em Adv. Theor. Math. Phys.} {\bf 2} (1998) 231--252,
  [\href{http://xxx.lanl.gov/abs/hep-th/9711200}{{\tt hep-th/9711200}}].

\bibitem{Gubser:1998bc}
S.~S. Gubser, I.~R. Klebanov, and A.~M. Polyakov, {\it Gauge theory correlators
  from non-critical string theory},  {\em Phys. Lett.} {\bf B428} (1998)
  105--114, [\href{http://xxx.lanl.gov/abs/hep-th/9802109}{{\tt
  hep-th/9802109}}].

\bibitem{Witten:1998qj}
E.~Witten, {\it Anti-de sitter space and holography},  {\em Adv. Theor. Math.
  Phys.} {\bf 2} (1998) 253--291,
  [\href{http://xxx.lanl.gov/abs/hep-th/9802150}{{\tt hep-th/9802150}}].

\bibitem{Aharony:1999ti}
O.~Aharony, S.~S. Gubser, J.~M. Maldacena, H.~Ooguri, and Y.~Oz, {\it Large n
  field theories, string theory and gravity},  {\em Phys. Rept.} {\bf 323}
  (2000) 183--386, [\href{http://xxx.lanl.gov/abs/hep-th/9905111}{{\tt
  hep-th/9905111}}].

  
\bibitem{'tHooft:1973jz}
  G.~'t Hooft,
   {\it A Planar Diagram Theory For Strong Interactions},
  %
  Nucl.\ Phys.\ B {\bf 72}, 461 (1974).


\bibitem{Holdom:1981rm}
B.~Holdom, {\it Raising the sideways scale},  {\em Phys. Rev.} {\bf D24} (1981)
  1441.

\bibitem{Holdom:1985sk}
B.~Holdom, {\it Techniodor},  {\em Phys. Lett.} {\bf B150} (1985) 301.

\bibitem{Yamawaki:1986zg}
K.~Yamawaki, M.~Bando, and K.-i. Matumoto, {\it Scale invariant technicolor
  model and a technidilaton},  {\em Phys. Rev. Lett.} {\bf 56} (1986) 1335.

\bibitem{Appelquist:1986an}
T.~W. Appelquist, D.~Karabali, and L.~C.~R. Wijewardhana, {\it Chiral
  hierarchies and the flavor changing neutral current problem in technicolor},
  {\em Phys. Rev. Lett.} {\bf 57} (1986) 957.

\bibitem{Appelquist:1987tr}
T.~Appelquist and L.~C.~R. Wijewardhana, {\it Chiral hierarchies and chiral
  perturbations in technicolor},  {\em Phys. Rev.} {\bf D35} (1987) 774.

\bibitem{Appelquist:1987fc}
T.~Appelquist and L.~C.~R. Wijewardhana, {\it Chiral hierarchies from slowly
  running couplings in technicolor theories},  {\em Phys. Rev.} {\bf D36}
  (1987) 568.
  


\bibitem{Manohar:1998xv}
  A.~V.~Manohar,
  arXiv:hep-ph/9802419.

\bibitem{Chivukula:1992gi}
  R.~S.~Chivukula, M.~J.~Dugan and M.~Golden,
  {\it Analyticity, crossing symmetry and the limits of chiral perturbation
   theory},
  %
  Phys.\ Rev.\ D {\bf 47}, 2930 (1993)
  [arXiv:hep-ph/9206222].


\bibitem{Tanabashi:1993sr}
  M.~Tanabashi,
   {\it Chiral perturbation to one loop including the rho meson},
  %
  Phys.\ Lett.\ B {\bf 316}, 534 (1993)
  [arXiv:hep-ph/9306237].


\bibitem{Georgi:1989xy}
  H.~Georgi,
   {\it Vector Realization Of Chiral Symmetry},
  %
  Nucl.\ Phys.\ B {\bf 331}, 311 (1990).



\bibitem{unpublished}
R.~S.~Chivukula, S.~Matsuzaki, E.~H.~Simmons, and M.~Tanabashi,
in progress.





\end{thebibliography}
\end{document}